\documentclass[12pt,preprint]{aastex}

\def\dw#1{}			
\def\jm#1{}

\newcommand{\op}{Ly$\alpha$\ }

\newcommand{\sdssA}{SDSSp J$143952.58-003359.2$~}
\newcommand{\sdssB}{SDSSp J$143951.60-003429.2$~}

\newcommand{\rxj}{RXJ0911.4+055}
\newcommand{\sdss}{Q1439-0034A,B}
\newcommand{\patnaik}{Q1422+2309A,Q1424+2255}
\newcommand{\pata}{Q1422+2309A}
\newcommand{\patb}{Q1424+2255}
\newcommand{\weed}{Q2345+007A,B}

\begin{document}


\title{
Expansion and Collapse in the Cosmic Web \altaffilmark{1,2}\\
 }

\vskip 1.5cm

\author{Michael Rauch\altaffilmark{3}, 
George D. Becker\altaffilmark{4}, Matteo Viel\altaffilmark{5}, Wallace L.W. Sargent\altaffilmark{4}, Alain Smette \altaffilmark{6,7},  Robert A. Simcoe\altaffilmark{8}, Thomas A. Barlow\altaffilmark{4},
Martin G. Haehnelt\altaffilmark{5}}
\altaffiltext{1}{Part of the observations were made at the W.M. Keck Observatory
which is operated as a scientific partnership between the California
Institute of Technology and the University of California; it was made
possible by the generous support of the W.M. Keck Foundation.}
\altaffiltext{2}{Based partly on observations collected at the European Southern Observatory, Paranal, Chile (ESO Programs 67.A-0371(A) and 69.A-0555(A).}
\altaffiltext{3}{Carnegie Observatories, 813 Santa Barbara Street,
Pasadena, CA 91101, USA}
\altaffiltext{4}{Astronomy Department, California Institute of Technology,
Pasadena, CA 91125, USA}
\altaffiltext{5}{Institute of Astronomy, Madingley Road, Cambridge CB3 0HA, U.K. }
\altaffiltext{6}{European Southern Observatory, Casilla 19001, Alonso de Cordova 3107, Santiago, Chile}
\altaffiltext{7}{Research Associate, F.N.R.S., Belgium}
\altaffiltext{8}{MIT Center for Space Research, 77 Massachusetts Ave., \#37-664B, Cambridge, MA
02139, USA}

\vfill

\begin{abstract}
We study the kinematics of the gaseous cosmic web at high redshift,
using Ly$\alpha$ forest absorption in multiple QSO sightlines.
Observations of the projected velocity shifts between Ly$\alpha$ absorbers
common to the lines of sight to a gravitationally lensed QSO
and  three more widely separated
QSO pairs are used to directly measure the expansion of the cosmic web in units
of the Hubble velocity, as a function of redshift and spatial scale. The lines of sight used span a redshift range from about 2 to 4.5
and represent transverse scales from the subkiloparsec range to about 300 $h^{-1}_{70}$ physical kpc.
Using a simple analytic
model and a cosmological hydrodynamic simulation we constrain the underlying three-dimensional distribution of expansion velocities from the observed line-of-sight distribution of velocity shear across the plane of the sky.  The shape of the shear distribution  and its width (14.9 kms$^{-1}$ rms for a physical transverse separation of
61 $h_{70}^{-1}$ kpc at z=2, 30.0 kms$^{-1}$ for 261 $h_{70}^{-1}$ kpc at z=3.6) are found to be in good agreement with 
the IGM undergoing large scale motions dominated by the Hubble flow, making this one of the most direct observations possible of the expansion of the universe. However, modeling
the Ly$\alpha$ clouds with a simple "expanding pancake" model, the average expansion velocity of
the gaseous structures causing the Ly$\alpha$ forest in the lower redshift ($z\sim 2$) smaller
separation (61 kpc) sample appears about 20\%  lower 
than the local Hubble expansion velocity.

In order to understand the observed velocity distribution further
we investigated the statistical distribution of expansion velocities in cosmological Ly$\alpha$ forest simulations. 
The mean expansion velocity in the ($z\sim 2$, separation $\sim$ 60 kpc) simulation is indeed  
somewhat smaller than the Hubble velocity, as found in the real data. 
We interpret this finding as tentative evidence for some Ly$\alpha$ forest clouds breaking away from the Hubble flow and undergoing the early stages of gravitational collapse. 
However, the distribution of velocities is highly skewed, and the majority  of Ly$\alpha$ forest clouds at all redshifts from 2 to 3.8  expand with
super-Hubble velocities, typically about 5\% - 20 \% faster than the Hubble flow. This
behavior is explained if most Ly$\alpha$ forest clouds in the column density range
typically detectable are expanding filaments that stretch and drain into more massive nodes.
The significant difference seen in the velocity distributions between the high and low 
redshift samples may conceivably reflect actual peculiar deceleration, the differences
in spatial scale, or our selecting
higher densities at lower redshift for a given detection threshold for Ly$\alpha$ forest lines.

We also investigate the alternative possibility
that the velocity structure of the  general Ly$\alpha$ forest could have an entirely different,  local origin as
expected if the Ly$\alpha$ forest were produced or at least significantly modified  by galactic feedback, e.g., winds 
from starforming galaxies at high redshift. However, we find no evidence that the observed distribution of velocity shear is significantly influenced by processes other than Hubble expansion and gravitational
instability. To avoid overly disturbing the IGM, galactic winds may  be  old and/or limp by the time we observe them in the Ly$\alpha$ forest, or they may occupy only an insignificant volume fraction of the IGM. We briefly discuss the observational evidence usually presented in favor of an  IGM afflicted by high redshift extragalactic superwinds and find much of it ambiguous. During the hierarchical
buildup of structure, galaxies are expected to spill parts of their interstellar medium and to heat and stir the IGM in ways that make it hard to disentangle this gravitational process from the effects
of winds.

\vspace{8.cm}
\end{abstract}


\clearpage

\section{Introduction}

Over the past decade our understanding of the general intergalactic medium (IGM), the  main baryonic component of the cosmic web,
has advanced considerably. Qualitative
questions
concerning the nature and interpretation of the IGM have given way increasingly to quantitative
investigations aimed at measuring astrophysical properties of the general
baryon field, among them the temperature, metallicity, kinematics, radiation field,
and dependence on  the underlying cosmological parameters.
More and more we are able to obtain distributions of the astrophysical
quantities as functions of time, spatial scale, and density, as opposed to mere mean values.

Most studies on the large-scale properties of the IGM so far have 
concentrated on the crucial problem of the physical scale of Ly$\alpha$ forest clouds.
The large sizes found (e.g., Weymann \& Foltz 1983; Foltz et al 1984; Smette et al 1992, 1995; Bechtold et al 1994; Dinshaw et al 1994, 1995;
Fang et al 1996; Crotts \& Fang 1998; Petitjean et al 1998;  Monier et al 1999; Lopez et al 2000; D'Odorico et al 1998, 2002; Williger et al 2000; Young et al 2001; Aracil et al 2002; Becker et al 2004)  have led to the realization that these clouds are really part of the general large scale
structure. Ionization arguments (Rauch \& Haehnelt 1995), analytical and Monte Carlo 
modelling of absorption in double lines of sight (Smette et al 1992, 1995; Charlton et al 1995;
Fang et al 1996; Crotts \& Fang 1998; Viel et al  2002) and cosmological hydro-simulations (Cen et al 1994; Petitjean et al  1995; Zhang et al 1995; Hernquist et al 1996; Miralda-Escud\'e et al 1996; Wadsley \& Bond 1997; Charlton et al 1997;
Cen \& Simcoe 1997) all suggest
that the absorbing structures are part of a filamentary cosmic web undergoing general
Hubble expansion, at least in an average sense.

\bigskip

In the present paper we argue that the observations of the velocity field in
the Ly$\alpha$ forest give us
insights into the earliest stages of structure formation, when overdense regions break away from the Hubble flow
and begin to collapse under the influence of gravity. 

We address the question as to how the gaseous cosmic web actually expands, as a function
of size, redshift, and density. We may reasonably expect that the cosmic web  should follow the Hubble flow on 
large (Mpc) scales, i.e. at least on scales larger than the typical coherence length of these
structures.
On intermediate scales (of order 100 kpc) the effects of gravitational collapse
may become more pronounced,  and galactic and sub-galactic potential
wells may impart kinetic energy to the gas, whereas on the smallest (subkiloparsec)
scales stellar evolution and gasdynamical processes in the interstellar medium
(ISM; supernova remnants, winds) must be the dominant sources
of kinetic energy and momentum.  
Earlier observations of small-scale structure in Ly$\alpha$ forest systems  have shown (Rauch et al 1999; 2001a,b) that there is also a trend
of the motions to increase in strength with increasing density, e.g., the higher
density gas appears to be more turbulent than the more typical Ly$\alpha$ forest cloud.

\medskip

To study  the velocity field of the IGM 
we exploit the fact that an anisotropic, randomly oriented, expanding gas cloud
will cause absorption features in two adjacent lines of sight intersecting it, that are shifted
relative to each other in proportion to the expansion velocity. Such shifts 
can be caused not only by the Hubble flow or gravitational collapse but by a wide range of
other processes including galactic feedback and systematic rotation.  Here we attempt to understand 
the origin of the observed motions. 

The paper is structured as follows.
The observations and data analysis are described in section 2, followed in section 3 by an analysis
of the velocity differences between common absorption systems at the smallest ($\sim 1$ kpc proper) scales as represented by the typical transverse separations between the beams to the
gravitationally lensed QSO RXJ0911.4+0551 (z=2.79; Bade et al. 1997). The cross-correlation
function between the two Ly$\alpha$ forest sightlines is derived, and an alternative method is presented that measures the difference between the line-of-sight velocities of individual, manually selected absorption systems common to two adjacent sightlines. The resultant distribution of velocity differences
for \rxj\ is discussed. Section 4 presents the same analysis at larger scales from
60 to about 300 $h_{70}^{-1}$kpc using the information from the QSO pairs Q2345+007A,B (z=2.16; Weedman et al 1982), Q1422+2309A (Patnaik et al 1992) and Q1424+2255 (z=3.63; Adelberger et al 2003), and \sdssA and \sdssB (hereafter abbreviated as Q1439-0034 A/B; z=4.25; Schneider et al 2000). The interpretation of the observed distribution is given first in light of a simple analytic model where
the Ly$\alpha$ clouds are randomly oriented, expanding pancakes of gas, in a discussion similar
to  Haehnelt (1996) and Charlton et al (1995, 1997). 
A further comparison is made with fake Ly$\alpha$ forest spectra from a cosmological hydrodynamic simulation (Viel et al 2004), which is analyzed
for velocity differences among absorbers in exactly the same way as the real data. 
Noting the excellent agreement for the distributions of velocity shear between data and simulation, we proceed to study and interpret the distribution of expansion velocities for absorbing clouds in the 
simulation. Section  5 on the possible influence of "cosmological superwinds" on
the kinematics of the IGM preceeds the final discussion and summary.

\section{Observations and Data Analysis of Multiple Lines of Sight}

\subsection{RXJ0911.4+0551}

\begin{figure}[p]
\includegraphics*[scale=0.65,angle=-90.]{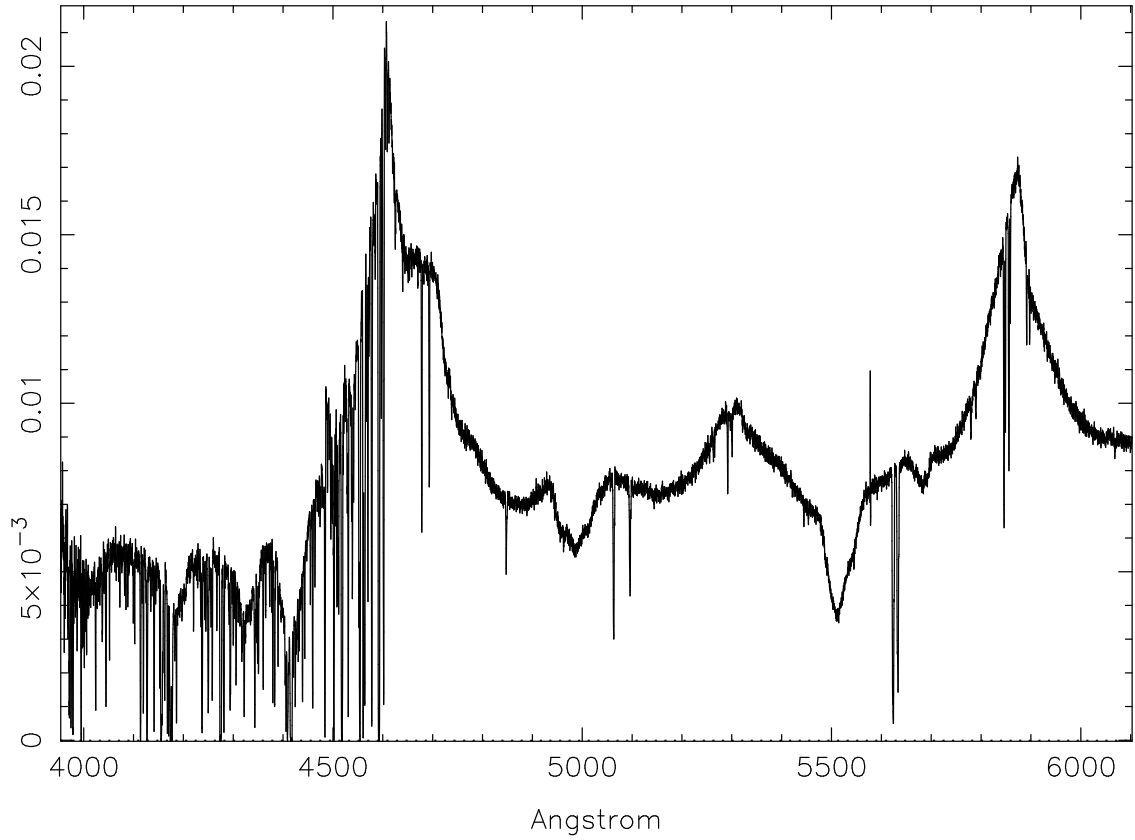}
\caption{\small ESI-spectrum of RXJ0911.4+0551AB. Note the broad absorption
troughs. The flux is in arbitrary units.\label{911spec}}
\end{figure}

RXJ0911.4+0551 is a radio-quiet z=2.79
QSO. This object was
identified as a gravitationally lensed quasar by Bade et al (1997). The
image configuration consists of four images in an "animal paw"
pattern with mutual separations of up to 3.1 arcsec (Burud et al
1998).  The object appears to be lensed by a galaxy cluster at z=0.7689
(Burud et al 1998; Kneib et al 2000).  The QSO itself is a
mini-broad absorption line (BAL) QSO  (Bade et al 1997); in the present data (see below) we detect triangular troughs at blueshifts of
18,700 kms$^{-1}$ relative to the QSO's broad CIV emission peak  (fig.\ref{911spec}). The absorption troughs are visible at z=2.559 in the transitions
HI 1216, NV 1239, 1243 \AA\ , CIV 1548,1551, SiIV 1393, 1403 \AA\ ,
and AlIII 1855, 1863 \AA . There is another  weaker trough at 5683 \AA
, probably another CIV component blueshifted by 10000 kms$^{-1}$ .

We observed the lensed images with the Keck II Echelle Spectrograph and
Imager (ESI; Sheinis et al 2002) on March 3 and 4, 2000, for 3600
seconds (A images) and 10800 seconds (B image). The spectra were taken
with a 0.75 arsec wide and  a 20 arcsec long slit leading to a spectral
resolution of 48.9 kms$^{-1}$ near the center of the spectrum.  The
slit was placed on the sky at a fixed position angle of 10 degrees.

Our results are based on a comparison between the spectra of a spatial
average of images A1,A2, and A3 on one hand and  image B on the other
(in the nomenclature for the lensed images established by Burud et al
1998). The B image was always well separated from the others, but
because of the small separation between the A images, it was not
possible to resolve them separately and components A1, A2, and A3 were
partly on the slit simultaneously.  We assume a nominal separation of
3.1 arcsec between the combined "A image" and B, and below we
refer to spectra A and B only, but one should keep in mind that the A
spectrum is a spatial average.

The data were extracted, wavelength-calibrated and fluxed using the
custom data reduction package
MAKEE\footnote{http://spider.ipac.caltech.edu/staff/tab/makee/index.html}
(Barlow \& Sargent 1997).  
The signal to noise ratios (S/Ns) in the Ly$\alpha$ forest region at 4260 \AA\
at the continuum  level are 56 and 27 per 3-pixel resolution element 
for the A and B images, respectively.

\begin{figure}[p]
\includegraphics*[scale=0.65,angle=-90.]{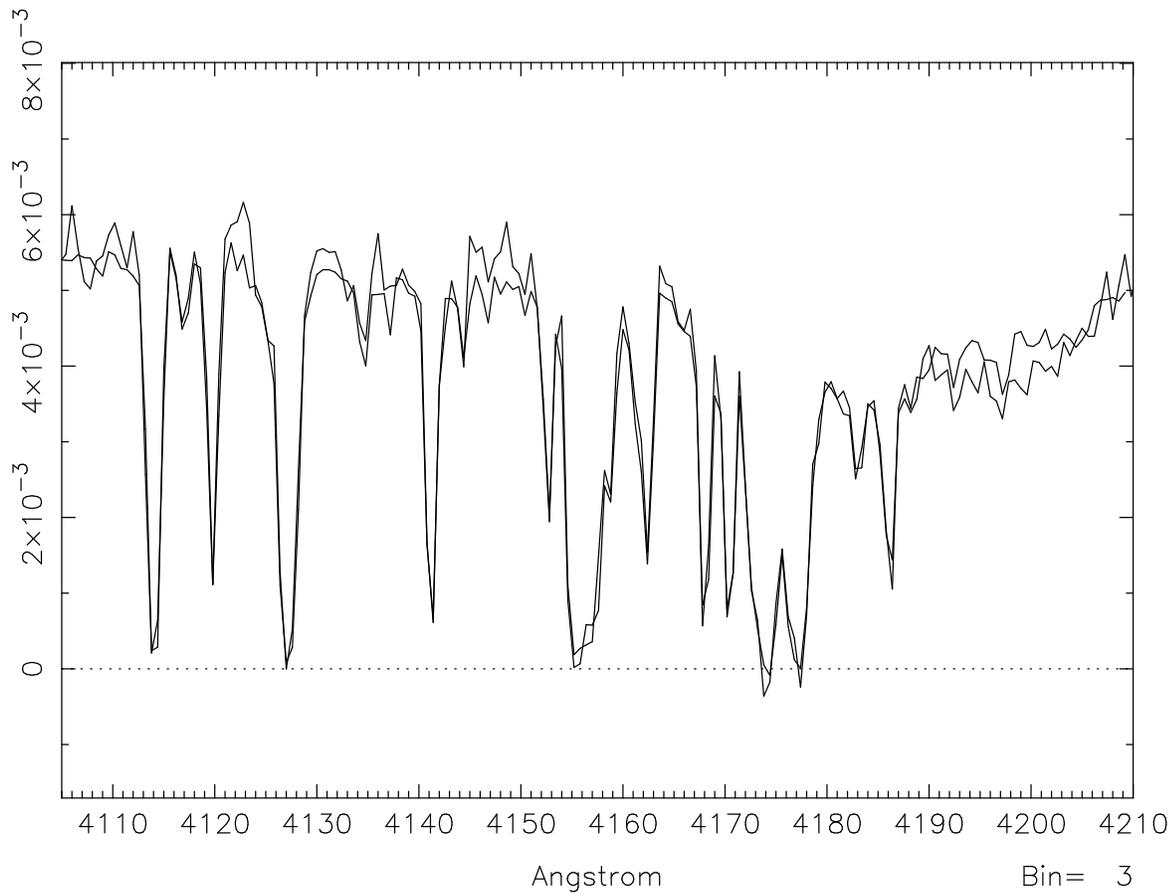}
\caption{\small Raw section of the spectra of RXJ0911.4+0551A,B prior to
flux calibration and continuum fitting. The flux is in arbitrary units.
The figure shows that the similarities between the Ly$\alpha$ forests are not
artifacts of the data reduction.\label{partspec}}
\end{figure}

To compare the absorption features between the
different spectra the overall shape of the spectra has to be matched.
The IRAF continuum task was used to fit a multiknot spline3 curve to
the ratio of the A and B spectra, ignoring spectral regions obviously
affected by absorption lines. The density of knots ranged from one
degree of freedom per 400 kms$^{-1}$ near the HI Ly$\beta$ emission
line to one per 340 kms$^{-1}$ near the HI Ly$\alpha$ line. This
approach wipes out genuine differences between the spectra on large
velocity scales but preserves differences between individual absorption
lines on scales smaller than a few hundred kms$^{-1}$. It also takes
out the BAL troughs. To illustrate
that the lack of differences between the spectra is not due mainly to
an overly flexible continuum fit we show in Fig. \ref{partspec} a section of the two
spectra (on top of each other) {\em before} any continuum fitting is
done. The mean proper transverse separations between the lines of sight
here is 1.1 h$_{70}^{-1}$ kpc. The spectra have only been scaled
globally to take out the overall difference in flux between the
images.  The similarity is remarkable and there are few obvious
differences in line strength and position for most individual
absorption lines. The section of the spectrum shown includes part of  one of the BAL troughs; there are some significant larger scale
variations between 4190 and 4210 \AA\ at low optical depths that are probably
caused by structure in the BAL outflow.

In any case,
it is clear just from visual inspection that the IGM
is highly homogeneous on kiloparsec scales.  Any differences in column density
and or velocity across the lines of sight must be subtle.

The following sections describe
various ways of quantifying this result.

\subsection{\weed}

This object, long suspected of being a gravitationally lensed QSO, 
has recently been shown (Smette et al 2005) to be a genuine QSO pair.
The data were obtained with the UVES instrument on the ESO VLT.
A total of 18600s over three exposures was obtained for image A and 60000s
over nine   exposures for  image  B.  All   observations were  carried in
service  mode between July 25, 2001  and  October 6, 2002 usually with
seeing conditions  better  than 0.8". The slit   was aligned along the
parallactic   angle  to  reduce  slit   loss  to  a minimum.   No  ADC
(atmospheric dispersion corrector) was used. The
data  reduction is described in A. Smette et al (2005, in preparation).

\subsection{\patnaik\ and \sdss}

The data and their reduction and a global correlation analysis of their 
Ly$\alpha$ forests are described in Becker et al (2004).
For the comparison between Q1422+2309 and Q1424+2255 only the spectrum of the "A" image of
Q1422+2309  was used.

\subsection{Contamination of the Ly$\alpha$ Forest by Metal Absorption Systems}

QSO metal absorption systems strong enough to be visible in spectra of
the current data quality (S/N ratios $\sim 10 - 70$) are usually
associated with strong, mostly saturated Ly$\alpha$ forest lines. As
shown in previous papers (e.g., Rauch et al 2001a), such "strong" metal absorbers
almost invariably show structure (velocity, column density changes)
over a few hundred parsecs. Thus, if the metal transition lines are
mistaken for HI Ly$\alpha$ lines, the turbulence of the IGM will be
overestimated and the correlation length of the IGM underestimated.  We have inspected the wavelength
stretches in the Ly$\alpha$ forest region potentially affected by
transitions belonging to metal absorbers identified from other
transitions redward of the Ly$\alpha$ line belonging to the same
system. Where the contamination was deemed significant these regions
were omitted from the analysis. Given our moderate S/Ns and resolution it is inevitable that some metal absorption systems are
being missed, especially if they have only lines embedded in
the Ly$\alpha$ forest.

\section{Searches for Structure in the Ly$\alpha$ Forest on Kpc Scales  }

This section discusses two methods to quantify differences between 
Ly$\alpha$ forest spectra from adjacent lines of sight: the cross-correlation function as a measure of 
{\it global} differences (section 3.1), and the comparison of the velocities of {\it individual} absorption
systems between the sightlines (section 3.2).

\subsection{Global differences between the spectra}

As in Rauch et al (2001b) and Becker et al (2004) we can study global differences in the \op forest region
by measuring the cross-correlation function $\xi_{cc}$ over the total useable length of both spectra
(see below).

We define this quantity again by 

\begin{eqnarray}
\xi_{cc} (\Delta v, \Delta r) \equiv
{ < (F_{\bf r}(v) - < F_{\bf r} > ) \cdot
    (F_{{\bf r}+\Delta {\bf r}}(v + \Delta v) - < F_{{\bf r}+\Delta {\bf r}} > ) > \over
   \sqrt{< (F_{\bf r}(v) - < F_{\bf r} >)^2 > \cdot
   <(F_{{\bf r}+\Delta {\bf r}}(v + \Delta v)-<F_{{\bf r}+\Delta {\bf r}}>)^2>} } ~. \label{eq:ccf} 
\end{eqnarray}

The quantities $F_{\bf r}$ and $F_{{\bf r}+\Delta {\bf r}}$ are the pixel flux
values of the two spectra, separated by $\Delta r$ on the plane of the
sky. The velocity coordinate along the line of sight is $v$ (where $dv
= d\lambda/\lambda$), and $\Delta v$ is the velocity lag. The averages
are taken over most of the velocity extent of the spectrum. For $\Delta
r$ = 0 we get the usual autocorrelation function $\xi_{cc}(\Delta v,
0)$, while for $\Delta v$ = 0 we have the cross-correlation as a
function of transverse separation\footnote{Throughout this paper beam
separations are computed for a flat universe with  $\Omega_m=0.25$ and $h_{70}=1$. In our earlier papers beam separations were given for a $\Omega_m=1.0,h_{50}=1$ cosmology, but the values differ by less than 20\% between the cosmological models, for the redshift range considered here.} only.  The function
is defined so as to satisfy $\xi_{cc} (0,0)=1$. With large-scale
velocity correlations ($>$ 1000 kms$^{-1}$) expected to be absent or
weak (Sargent et al. 1980), the autocorrelation function (on scales of 
$\sim$ 100 kms$^{-1}$) mostly measures the \op line width and the weak
small-scale clustering of \op forest systems (e.g., Webb 1987; Rauch et
al.  1992).  We apply the correlation analysis to the wavelength
interval [3950,4614] \AA\ of 
the QSO in our sample with the smallest separation 
between its images, RXJ0911.4+0551A,B. Thus most of the spectral
region between Ly$\beta$ and Ly$\alpha$ emission is included,  omitting only
a small  region [4381,4386] \AA\ where there is a significant contamination by
a known metal SiIII 1206 \AA\ interloper at z=2.633.   The resulting  mean redshift $\overline{z}$ =
2.522 of the remaining sample corresponds to a mean beam separation\footnote{The redshift of
the lens of \rxj\ is taken to be z = 0.7689 (Kneib et al 2000).}
$\overline{\Delta r}$ = 1.0 h$^{-1}_{70}$kpc.  

\begin{figure}[p]
\includegraphics*[scale=0.65,angle=-90.]{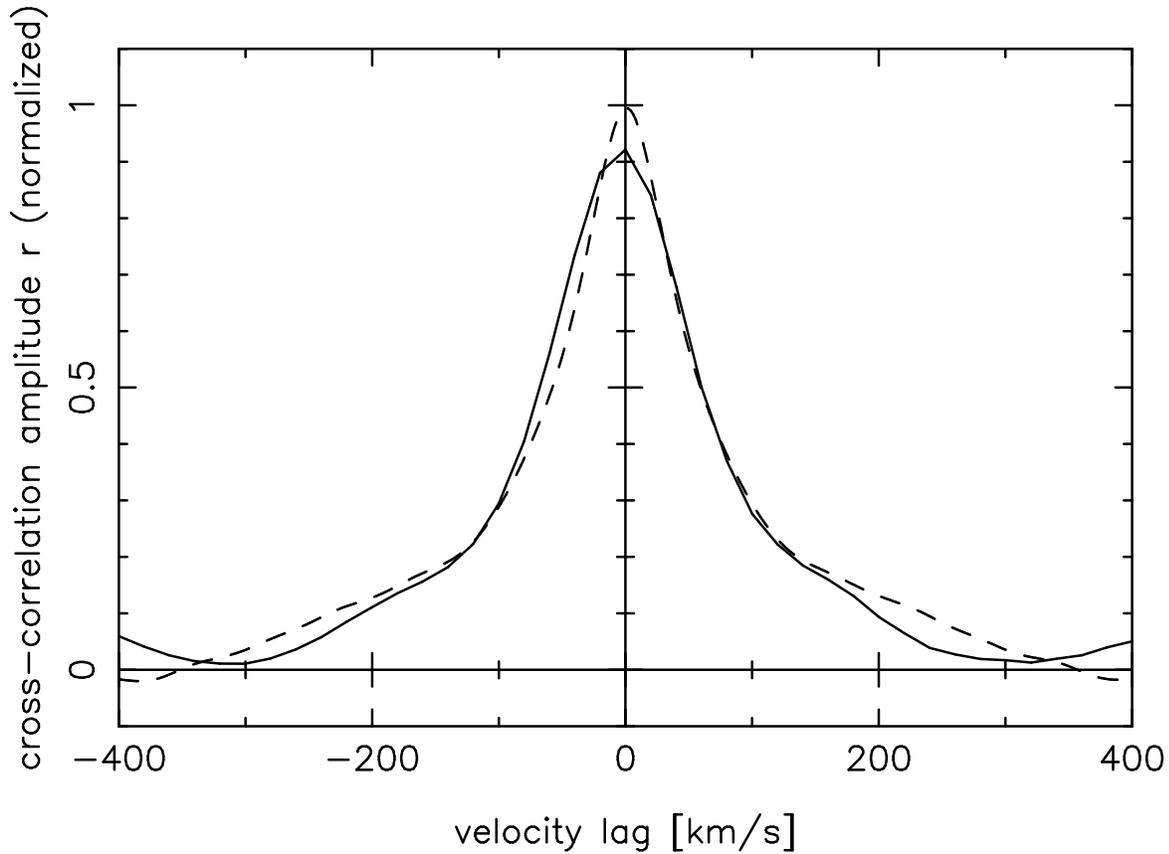}
\caption{\small
cross-correlation function $r(<d>,\Delta v$), for a mean beam
separation $<d>$ = 1.0\ $h_{70}^{-1}$\ {\em kpc}, between the \op
forests in the RXJ0911.04+0551 A and B (solid line). For comparison,
the same function is shown for the closer separation  ($<d>$ = 0.108\ $h_{70}^{-1}$\ {\em kpc}) Q1422+231 A and C image pair (dashed line). The peak for the RXJ0911 case
is slightly shifted to the left because of uncertainties in the placement of the
images on the spectrograph slit (see text). \label{crosscomparis}}
\end{figure}

The function 
$\xi_{cc} (\Delta v,\overline{\Delta r})$ is shown in
fig. \ref{crosscomparis}.  In particular, we obtain the "zero-lag"
cross-correlation function for the RXJ0911.4+0051 lines of sight:
\begin{eqnarray}
\xi_{cc}(\Delta v = 0;\ \overline{\Delta r} = 
1.0\ h_{70}^{-1}\ {\rm kpc})=\ 92.1\ \% .
\end{eqnarray}

For comparison, the dashed line shows the same quantity for the pair
of sightlines between the Q1422+231 images investigated in Rauch et al 2001b.
The mean separation there is about an order of magnitude smaller:
\begin{eqnarray}
\xi_{cc}(\Delta v = 0;\ \overline{\Delta r} = 
108\ h_{70}^{-1}\ {\rm pc})=\ 99.5\ \% .
\end{eqnarray}

Thus, even at the larger kiloparsec separation probed with the new, RXJ0911.4+0551 data,
the global differences between the spectra are very small, indicating that
the average coherence length in the IGM is  much larger
than a kiloparsec.

\subsection{Local Differences: Velocity Shear in Individual Absorption Systems}

The above correlation analysis has only shown that the cosmic web {\it on average} is highly coherent
on kiloparsec scales. Nevertheless, infrequent but strong local differences in  column density as well as velocity shifts (caused by galactic winds, rotation, or any
small-scale structure in the ISM of an intersected galaxy) could
easily manifest themselves
on scales of a few hundred km s$^{-1}$ without degrading the cross-correlation signal
significantly. 

To investigate this possibility and to get a more quantitative understanding of
what is happening at the level of a single absorbing cloud, we have searched for velocity shifts among {\it individual}
absorption lines or small complexes  between the two lines of sight.
We had attacked this question previously in Rauch et al (2001b), where the Ly$\alpha$
forest lines in Q1422+231 had been modeled with Voigt profiles. The
decomposition into multiple Voigt profiles becomes more ambiguous 
at the lower (ESI) spectral resolution available here, which makes the pairwise comparison
between components in separate lines of sight less certain.

\begin{figure}[p]
\includegraphics*[scale=0.65,angle=-90.]{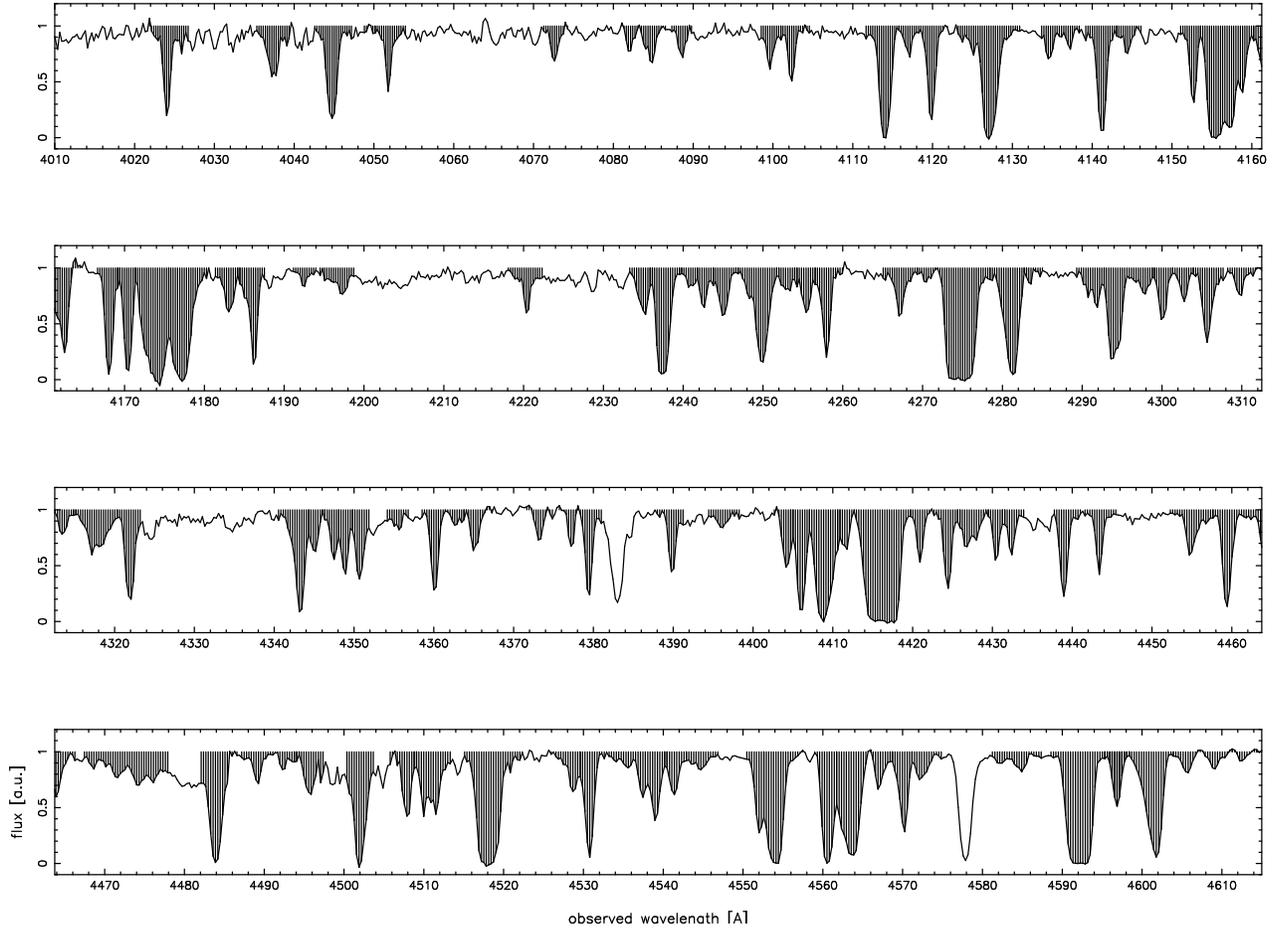}
\caption{\small Ly$\alpha$ forest spectrum of RXJ0911.04+0551A with the filled parts indicating the
regions used for the measurement of the velocity differences. The omitted
regions were either deemed to depart too little from the continuum 
or were affected by metal line interlopers.\label{regions}}
\end{figure}

Thus, in the present case a different, less model-dependent method was adopted. Individual
absorption lines are  selected by eye, by marking a wavelength window including the line 
with a cursor. It was generally attempted to delineate the absorption lines by marking the points on either
side of the line center where the continuum  had substantially recovered, but this approach was not always possible
and sometimes a much closer section around the line center had to be chosen to avoid
contributions from a blended component with seemingly different kinematics. However, the measurement
should not be very sensitive to the exact width and position of the window, as this
is a relative measurement and it is only important that the same window be imposed on both spectra.
This is repeated for all lines deemed to be HI Ly$\alpha$.  Then the difference 
between the flux-weighted projected velocities, or the {\it velocity shear}, 
$\Delta\overline{v} = \overline{v_B} - \overline{v_A}$, is computed for each window along the 
lines of sight A and B, where
the velocity weighted by the absorbed flux is defined as
\begin{eqnarray}
\overline{v} = \frac{\sum_i w_i v_i }{\sum_j w_j}.\label{vmean}
\end{eqnarray}
Here the flux weight $w_i$ of a pixel $i$ is
$w_i=(1-f_i)$, with $f_i$ being the flux relative to a unit continuum,
and the summation is over all pixels of the chosen 
spectral window enclosing the absorption line. The width of the pixels is constant
in velocity space. The origin of the velocity coordinate is defined to be
the center of the window around the absorber.
The absorption regions of the spectrum included in the analysis are shown as blackened 
areas in fig. \ref{regions}.

To see whether there are intrinsic velocity shifts $\Delta\overline{v}$ between the lines of sight exceeding the scatter due to measurement uncertainties, the variance in the velocity measurement needs to be computed.

The variance in the determination of $\overline{v}$ from eqn. \ref{vmean} is then
\begin{eqnarray}
\sigma^2(\overline{v}) = \frac{\sum_i v_i^2 \sigma^2(w_i) }{\left(\sum_j w_j\right)^2}+
\frac{\left(\sum_k w_k v_k\right)^2\sum_i \sigma^2(w_i)}{\left(\sum_j w_j\right)^4}, \label{singvar}
\end{eqnarray}
where $\sigma(w_i)$ is the standard deviation of the normalized absorbed flux
of pixel $i$. This is just the error of the flux in that pixel
as derived from the original error array, based on photon counting
statistics. 
There is no term accounting for the error in the velocity calibration,
which we assume to be negligible, for the time being.

The variance in the velocity difference is simply
\begin{eqnarray}
\sigma^2(\Delta\overline{v}) = \sigma^2(\overline{v_B}) + \sigma^2(\overline{v_A}). \label{varvar}
\end{eqnarray}

\begin{figure}[p]
\includegraphics*[scale=0.65,angle=-90.]{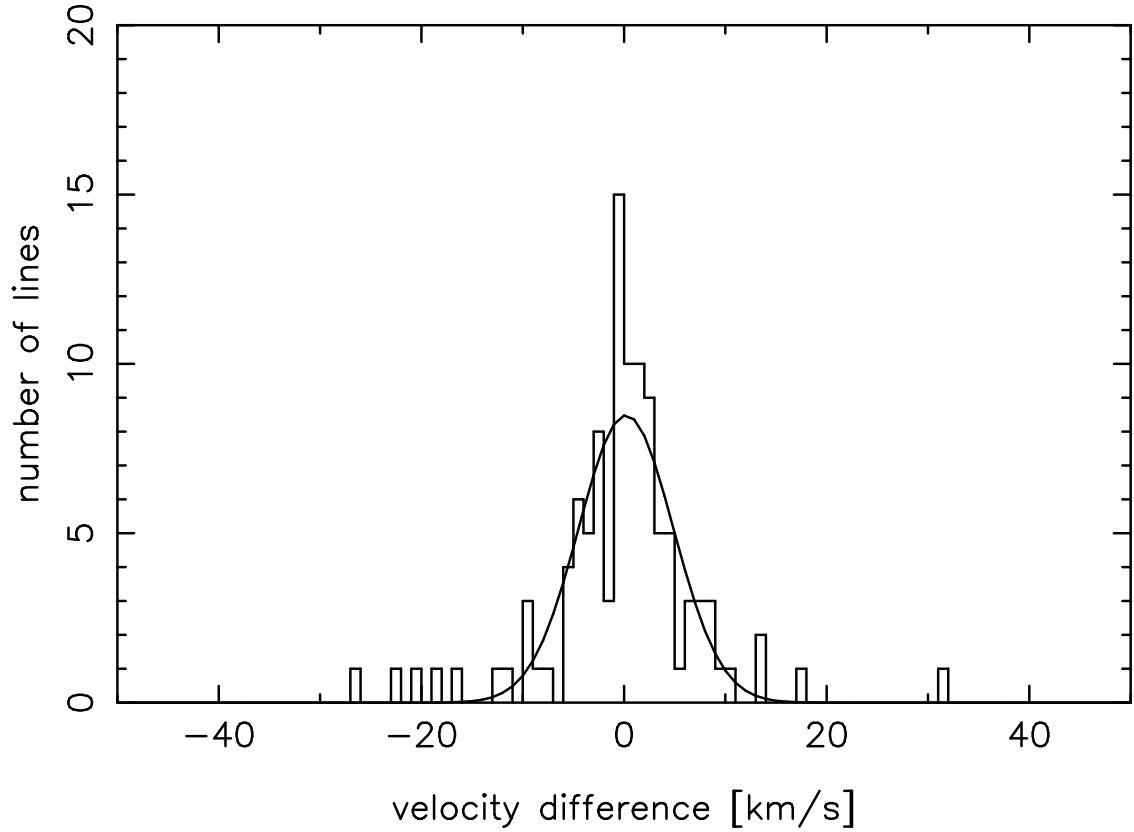}
\caption{\small Observed histogram of the velocity differences $\Delta v$ (=
$v_B$ -- $v_A$) for 
pairs of absorption components toward RXJ0911.04+0551A,B. The solid curve is the expected
Gaussian distribution, if the width were entirely caused by the mean measurement error $\sigma(v_B-v_A)=4.7$ kms$^{-1}$, and there were no intrinsic differences between the lines of sight.
There are a few outliers with $3\sigma$ significant velocity differences that are shown
individually in fig.\ref{3sigdevplot}.  \label{vdiffs} }
\end{figure}

The distribution of the observed differences in projected velocities
between the lines of sight, $\Delta\overline{v}$, is shown in Fig. \ref{vdiffs}.
The data are taken from 108 absorption regions spanning 4000 - 4614 \AA
, i.e., from Ly$\alpha$ not quite down to Ly$\beta$ (a noisy bit at the
short wavelength end below 4000 \AA\ was left out). Ly$\alpha$ lines
in four other regions were left out because of blends with metal-line
interlopers from an absorption system at higher redshift:  regions 4383
-- 4386 \AA\ and 4323--4339 \AA\ were affected by blends with the SiIII
1207 \AA , and SiII 1190,1193 \AA\ lines, respectively, from a system
at $z=2.6327$. Similarly, the Ly$\alpha$ line near 4577.9 \AA\ is
blended with SiII 1260 \AA\ from the same absorber.

The mean velocity shift between the lines of sight, obtained from the
average of all velocity shifts of all remaining 108 regions, weighted
by the inverse of the square of the  measurement error, was found to be
$1.63 \pm 0.17$ kms$^{-1}$. Such a shift has been seen before between
lensed spectra of QSOs (Rauch et al 2001b) and is most likely caused by the
uncertainty involved in placing both images (sequentially) at the
same position on the spectrograph slit. When comparing the actual
distribution of the velocity differences with the one predicted by
observational scatter, the mean shift was subtracted first.

The 108 lines are at mean redshift
2.567, corresponding to a transverse separation of 0.82 $h^{-1}_{70}$ kpc between the beams.

The {\it observed} absolute value of the velocity difference, averaged
over all regions, amounts to

 \begin{eqnarray}
<|\Delta\overline{v}|> = 4.9\ {\mathrm kms}^{-1}.
\end{eqnarray}

The observed rms velocity difference is 

\begin{eqnarray}
\sqrt{<\Delta^2\overline{v}>}= 7.4 \ {\mathrm kms}^{-1}, 
\end{eqnarray} 
whereas the  standard deviation for the velocity differences
predicted on the basis of the measurement errors alone is

\begin{eqnarray}
\sigma(\Delta\overline{v})= 4.7\ {\mathrm kms}^{-1.} \label{eq:rms}\end{eqnarray}
A $\chi^2$ test
shows that the innermost $\pm 13$ kms$^{-1}$ (equivalent to 2.8 $\sigma$) of the observed distribution of projected velocity differences
between the lines of sight, $\Delta\overline{v}$,  
has a 40\% probability of having been drawn from a Gaussian error distribution with
$\sigma(\Delta\overline{v}) = 4.7$ kms$^{-1}$ (Fig. \ref{vdiffs}), i.e., most of the velocity differences are  consistent
with pure measurement error.

However, there are hints of some significant excursions beyond mere measurement uncertainty.
Of the observed velocity
differences, 37\%  exceed 1 $\sigma$ 
if predicted by the measurement
error, and  4.6\% (five systems) are beyond 3 $\sigma$ (eqn. \ref{varvar}). Note that the excursions here
are with respect to the individual measurement uncertainty for a particular region, which
generally is different from the width of the distribution, eq.[\ref{eq:rms}].

\begin{figure}[p]
\includegraphics*[scale=.8,angle=-0.]{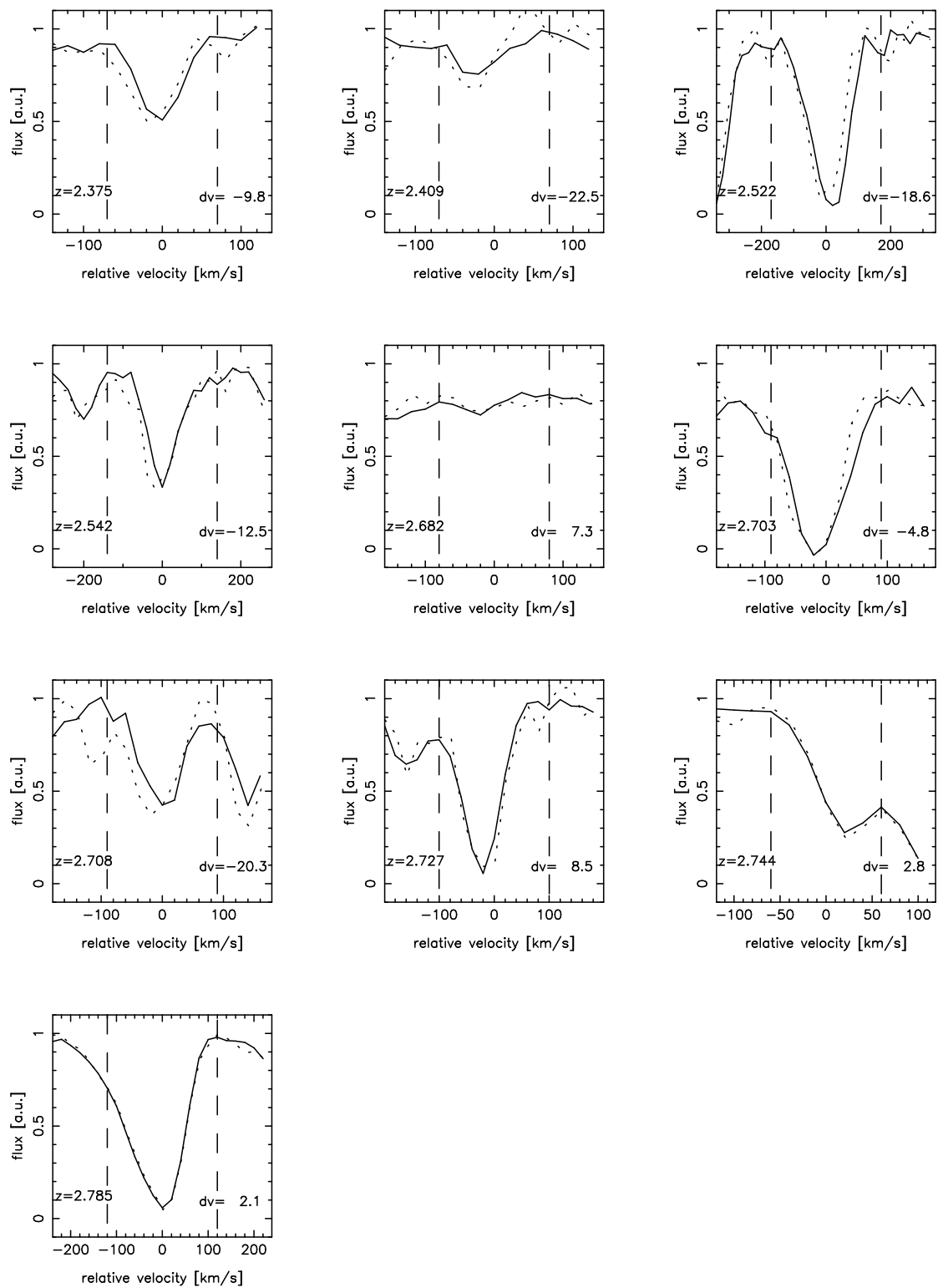}
\caption{\small Absorption lines toward RXJ0911.04+0551A,B with velocity differences between
the lines of sights larger than 2.5 standard deviations. The differences between the spectra
appear to be mostly consistent with velocity shifts of the entire absorption system.
The measured velocity shifts $dv$ are shown in each panel in units of kms$^{-1}$.\label{3sigdevplot}}
\end{figure}

The 10 cases of absorption lines with larger than 2.5 $\sigma$ velocity shifts
are shown in fig. \ref{3sigdevplot}. Perhaps half of them are borderline cases
where a bad continuum fit or some defect in one spectrum could have produced an
artifact. None of the remaining systems exhibit any unusual evidence for
strong turbulence or strong column density gradients, but they appear to be consistent
with a mere {\it velocity shift of the entire absorption system}. The mean absolute shift
in these 10 cases is 11 kms$^{-1}$.

Subtracting in the above cases the predicted width of the distribution based on errors alone from the measured rms width in quadrature we need to explain an additional width of about 6 kms$^{-1}$ rms as having a physical origin. We can only
speculate about the origin of these shifts. 
The Hubble expansion over kiloparsec distances
like the ones considered here would only cause immeasurably small velocity gradients. The most likely
explanation appears to be the presence of a nearby gravitational potential well (associated
with the grainy mass distribution in the filaments), in which
the gas is "circling the drain", i.e., undergoing rotation or differential motion during gravitational infall.

\section{The Transition to Larger Scales}

\begin{figure}[p]
\includegraphics*[scale=0.65,angle=-90.]{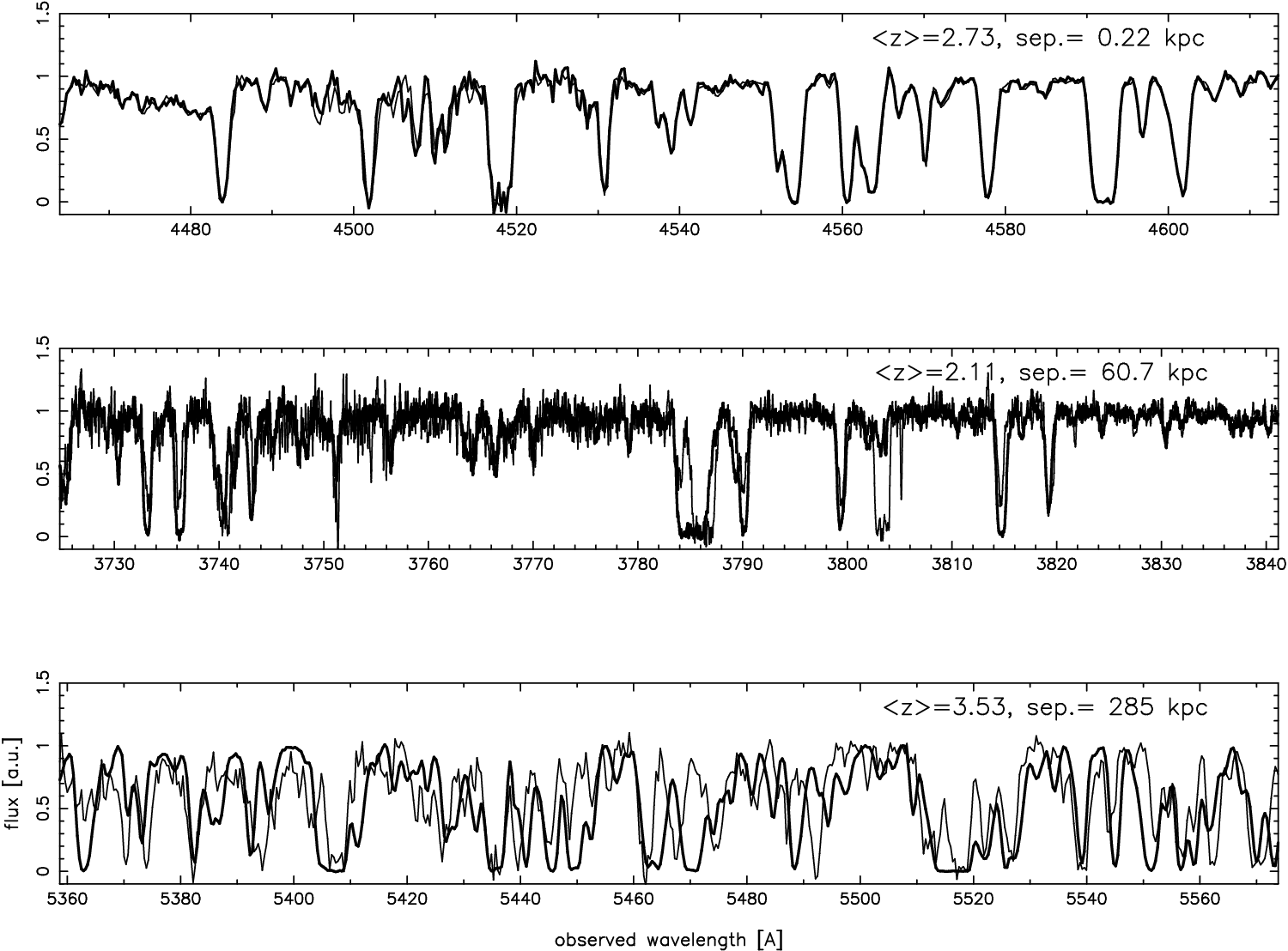}
\caption{\small 
Sections of the three Ly$\alpha$ forest double lines of
sight, in the order of increasing separation between the lines of sight (from top to bottom: \rxj, \weed,  \patnaik). The length of the spectra
is chosen in all cases to be 100 $h^{-1}$ comoving Mpc. The mean redshifts
and the mean beam separation (in physical $h^{-1}_{70}$ kpc) are given in the right upper corner of the spectra. 
 The 
discrepancies between the column densities and velocities of the individual absorption lines are generally insignificant for the case
with sub-kpc beam separation, but they become noticeable at 60 kpc and quite dramatic at 285 kpc. Note that even in the last case there still is quite 
a bit of similarity between the lines of sight.\label{comparisspec}}
\end{figure}

With increasing transverse separation between the
lines of sight absorption systems become increasingly decoherent, as can be
seen from a comparison of sections of the spectra of \rxj, \weed, 
and \pata, \patb\   (fig.\ref{comparisspec}). The figure shows three sections of the QSO lines of sight chosen to have equal
comoving extent of 100 $h^{-1}$ Mpc. The spectra along the two lines of sight to two different QSO images in each panel are represented by a thick line and a thin line.
The spectra  to \rxj\ are  essentially identical
in both lines of sight, over a mean transverse separation of 0.22 $h^{-1}_{70}$ physical kpc.
Aside from the obvious differences in S/N and mean absorption (note the 
different redshifts between the panels), the most obvious change when going from the top to the bottom panel is the increasing dissimilarity between the spectral pairs.

 The \weed\ spectra at a mean separation of 60.7 $h^{-1}_{70}$ proper kpc already differ somewhat in 
the column densities and positions of individual lines, but all of the systems can still
easily be cross-identified among the lines of sight. For the case of \pata, \patb\  shown
here (from Becker et al 2004), at a mean separation of 285 $h^{-1}_{70}$ kpc there are strong differences for individual systems, which often cannot be traced easily across the lines of sight.
Nevertheless, voids (regions of low absorption) and strong lines can still be recognized
reasonably often  across the lines of sight if one allows for some shifts in the projected velocity and for
column density differences.

Clearly there are scales where the  observed velocity shear (i.e., the differences between
the velocities projected along the line of sight of two absorption lines observed in 
adjacent lines of sight) must be dominated by the underlying systematic expansion of the cosmic web.
With the exception of the case of \rxj, the beam separations for the QSOs considered here
are large enough that a significant amount of the velocity shifts between individual
absorption lines across the lines of sight should arise in the Hubble flow.

\subsection{The Observed Distribution of Velocity Shear}

\begin{figure}[p]
\includegraphics*[scale=0.65,angle=-90.]{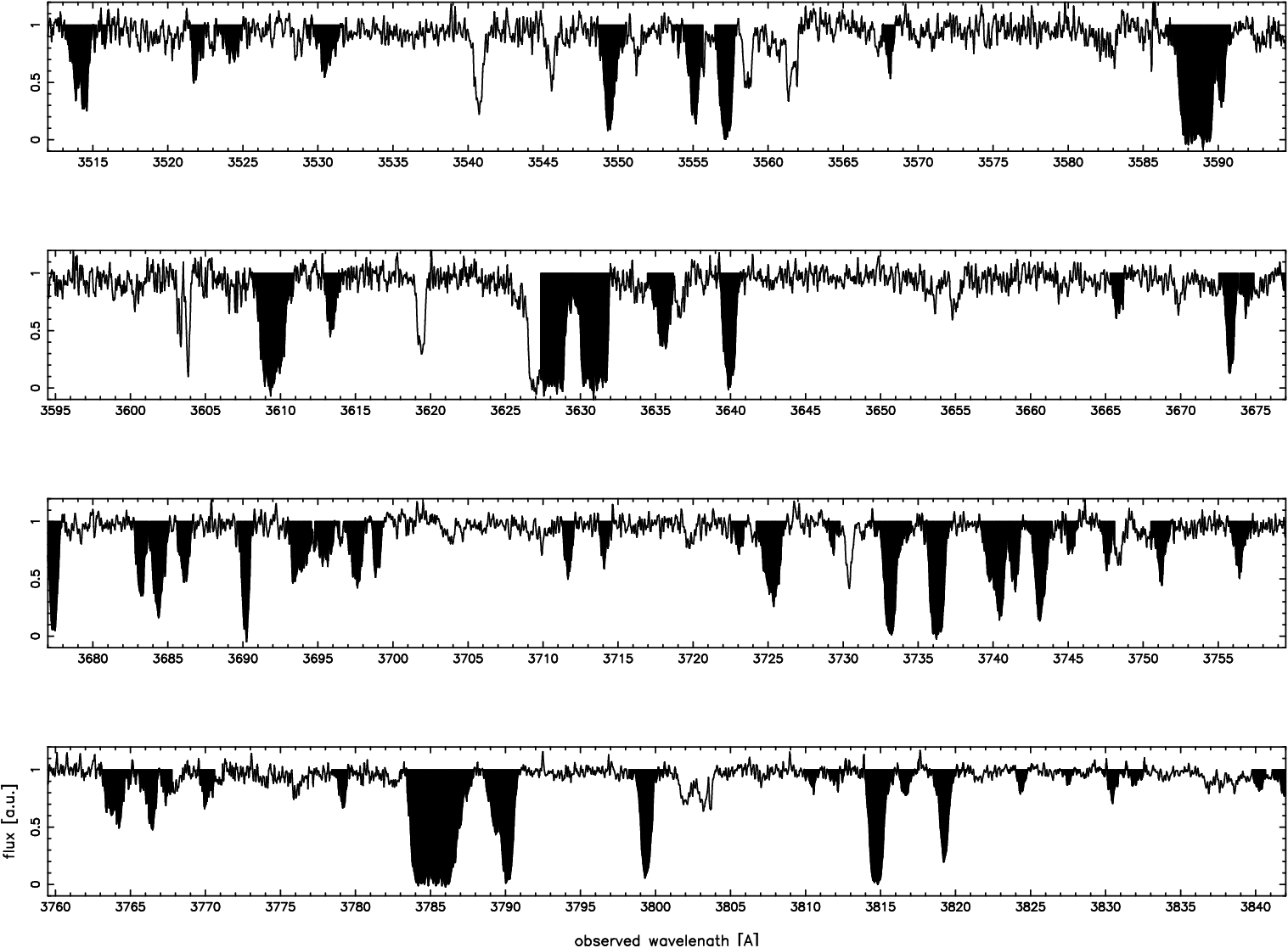}
\caption{\small Ly$\alpha$ forest spectrum of \weed\ with the filled parts indicating the
regions used for the measurement of the velocity differences. The omitted
regions were either deemed to depart too little from the continuum 
or were affected by metal line interlopers.\label{2345regions}}
\end{figure}

\begin{figure}[p]
\includegraphics*[scale=0.65,angle=-90.]{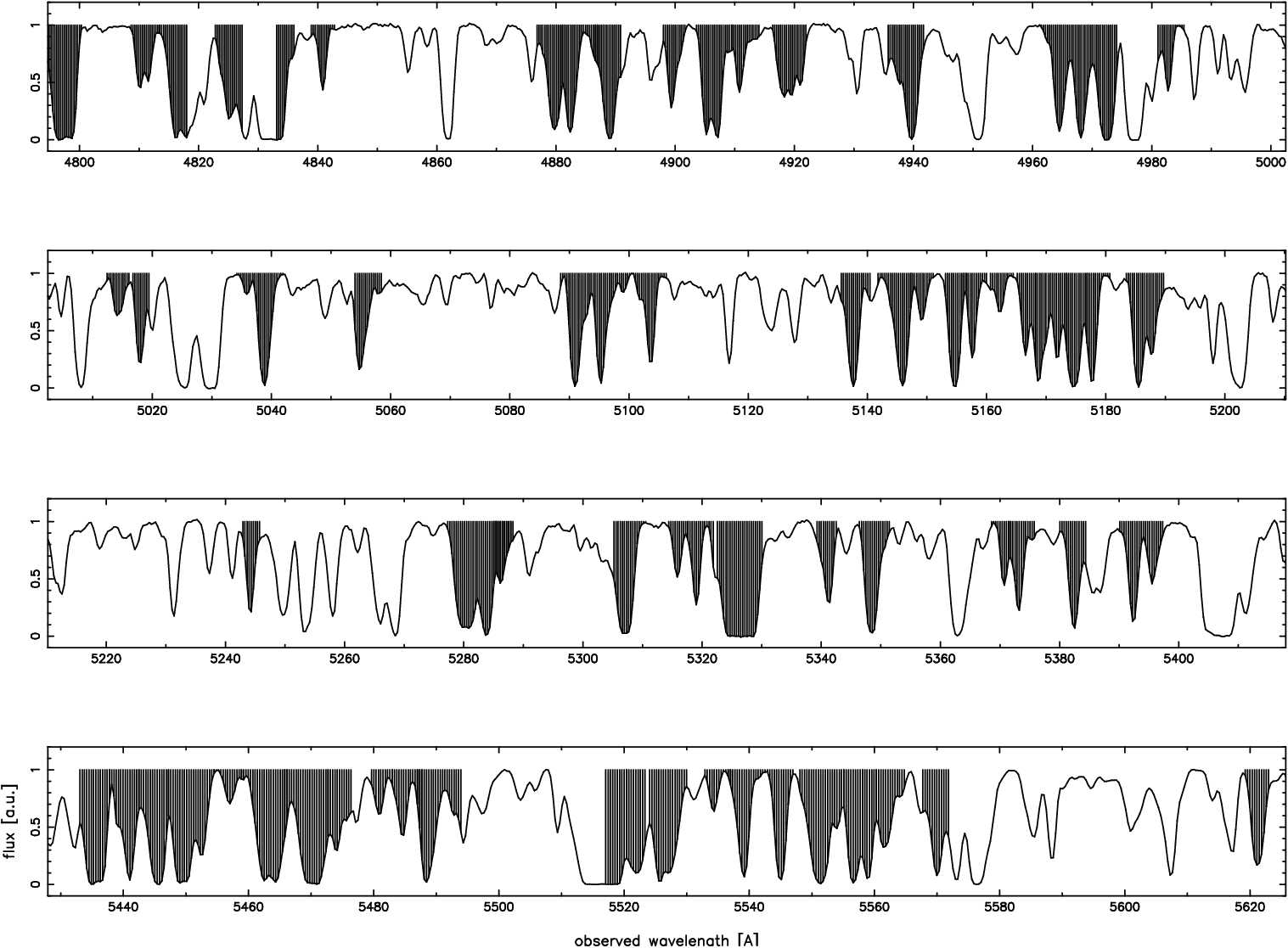}
\caption{\small Ly$\alpha$ forest spectrum of \pata, \patb\  with the filled parts indicating the
regions used for the measurement of the velocity differences.\label{1422regions} }
\end{figure}

\begin{figure}[p]
\includegraphics*[scale=0.65,angle=-90.]{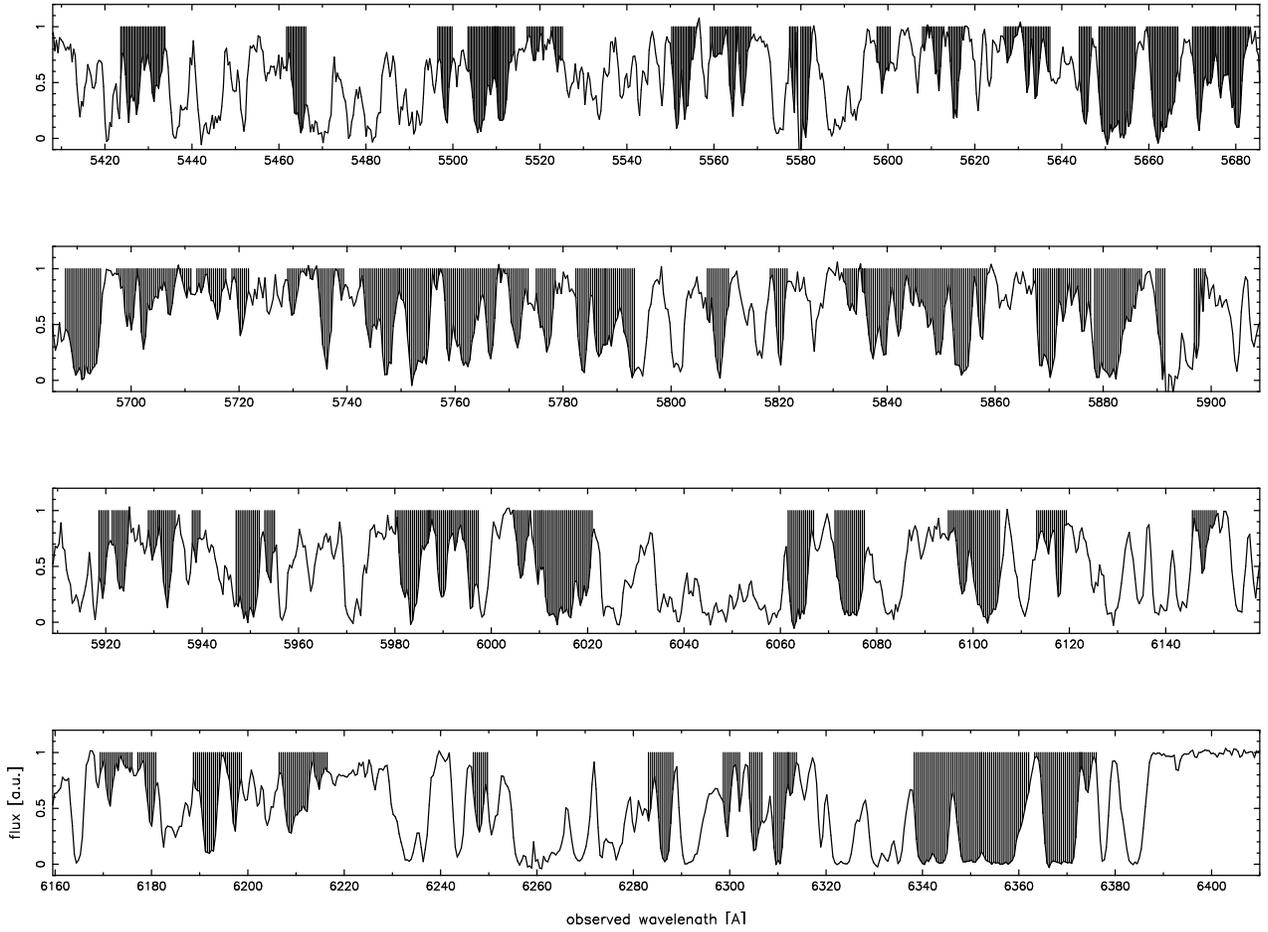}
\caption{\small Ly$\alpha$ forest spectrum of \sdss\  with the filled parts indicating the
regions used for the measurement of the velocity differences.\label{1439regions} }
\end{figure}

\begin{figure}[p]
\includegraphics*[scale=.6,angle=-90.]{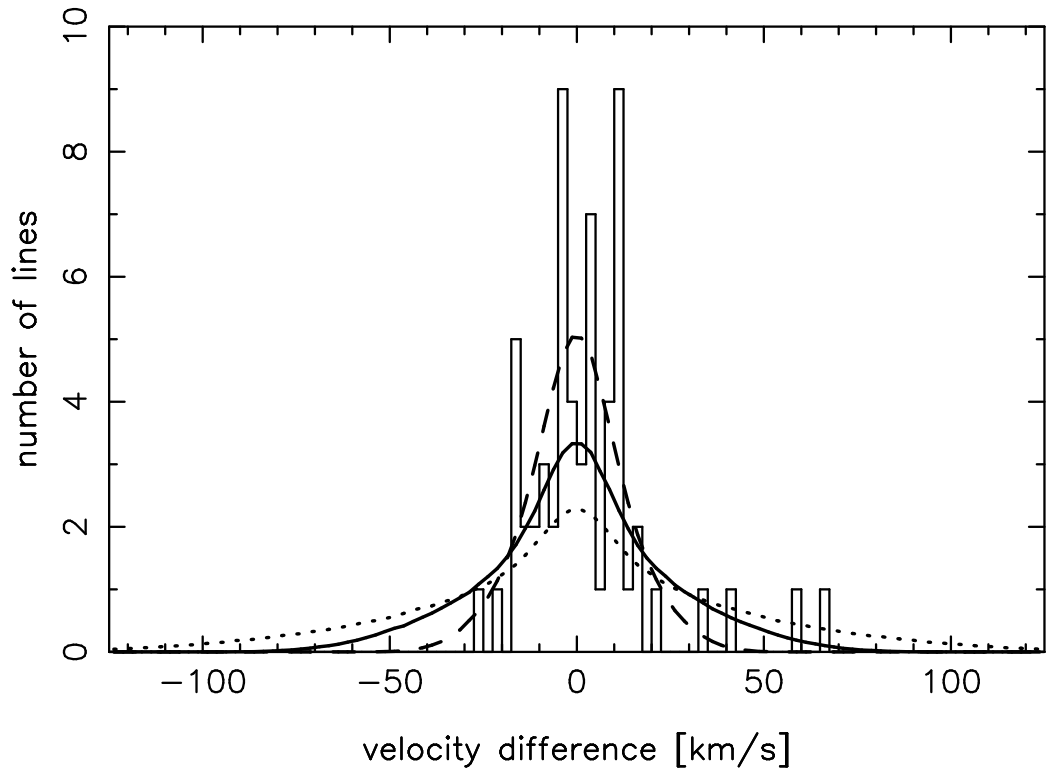}
\caption{\small velocity shear between the lines of sight for 
Ly$\alpha$ forest lines in the spectra of Q2345+005A and B.
The histogram gives the {\it observed} distribution of the measured shear between
corresponding absorption lines in the two lines of sight. The solid line is the best fit expanding pancake model with $v = 0.8\times v_{\mathrm Hubble}$. For comparison, the dashed (dotted) lines show the model distribution
for  $v = 0.4 (1.5)\times  v_{\mathrm Hubble}$, respectively.\label{q2345_velewmodels} }
\end{figure}

\begin{figure}[p]
\includegraphics*[scale=.6,angle=-90.]{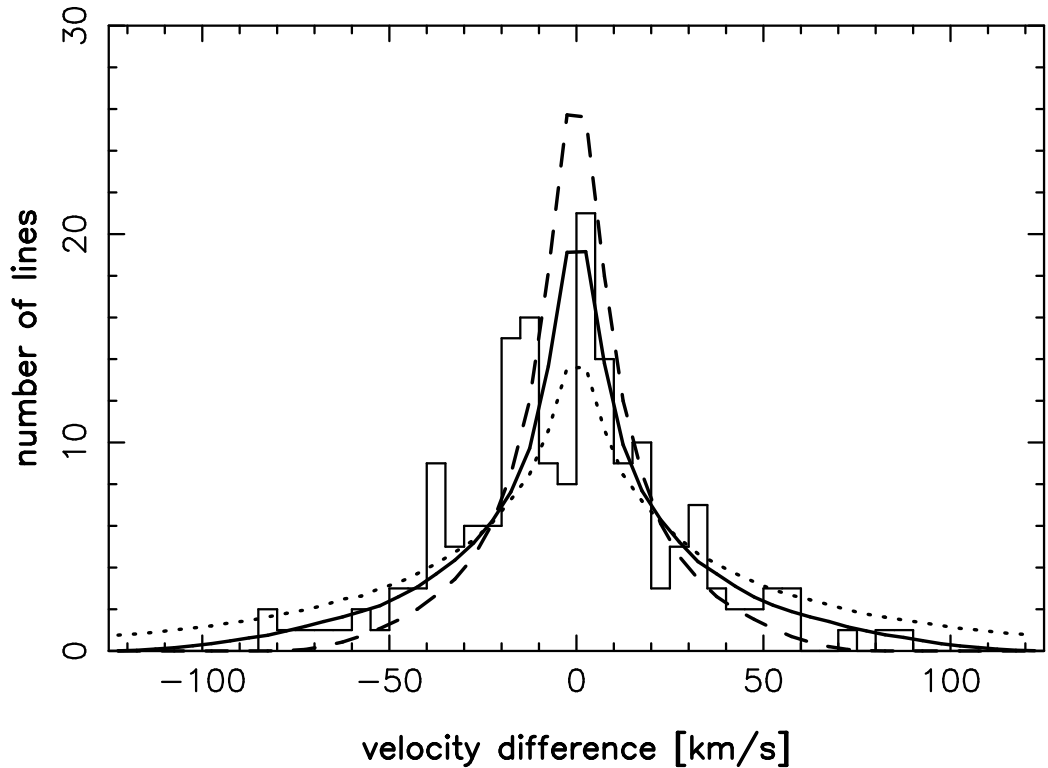}
\caption{\small velocity shear between the lines of sight for 
Ly$\alpha$ forest lines in the spectra of Q1439-0034A and B,
and 
the pair Q1422+2309A and Q1422+2255.
The histogram gives the combined {\it observed} distributions of the two pairs. The solid line is the best fit expanding pancake model with $v = 0.65\times v_{\mathrm Hubble}$. For comparison, the dashed (dotted) lines show the distribution
for  $v = 0.4 (1.2)\times  v_{\mathrm Hubble}$, respectively. \label{beckervelewmodels}}
\end{figure}

The transition to larger scales dominated by the Hubble flow should be visible
as a change in shape of the distribution function of velocity shear. For the wider separation pairs  the flux-width weighted velocity differences between the lines of sight  were measured as before, for individual
absorption systems in regions selected by eye. 
A uniform minimum rest frame equivalent width of 0.4 $\AA\ $ was required  as a necessary but not  sufficient condition for all lines in
order to be included in the samples.
Because of the wider separations 
not all systems could be successfully cross-identified.  Doubtful cases, where the
continuation of an absorption system across the sky was ambiguous, were omitted,
leading us to err on the conservative side.
Figs. \ref{2345regions},\ref{1422regions} and \ref{1439regions}, show again the selected regions as "blackened" and give an illustration of the severity of this selection effect. The observed
distributions for the wider separation QSO pairs are shown as histograms in figs. \ref{q2345_velewmodels} (for \weed), and \ref{beckervelewmodels} (where the velocity shear measurements of \pata\ and \patb\ and of 
\sdss\ have been combined into one histogram because the redshifts and separations are similar). The velocity shifts were determined in the same way as described for \rxj\ 
above.

\begin{figure}[p]
\includegraphics*[scale=.6,angle=-90.]{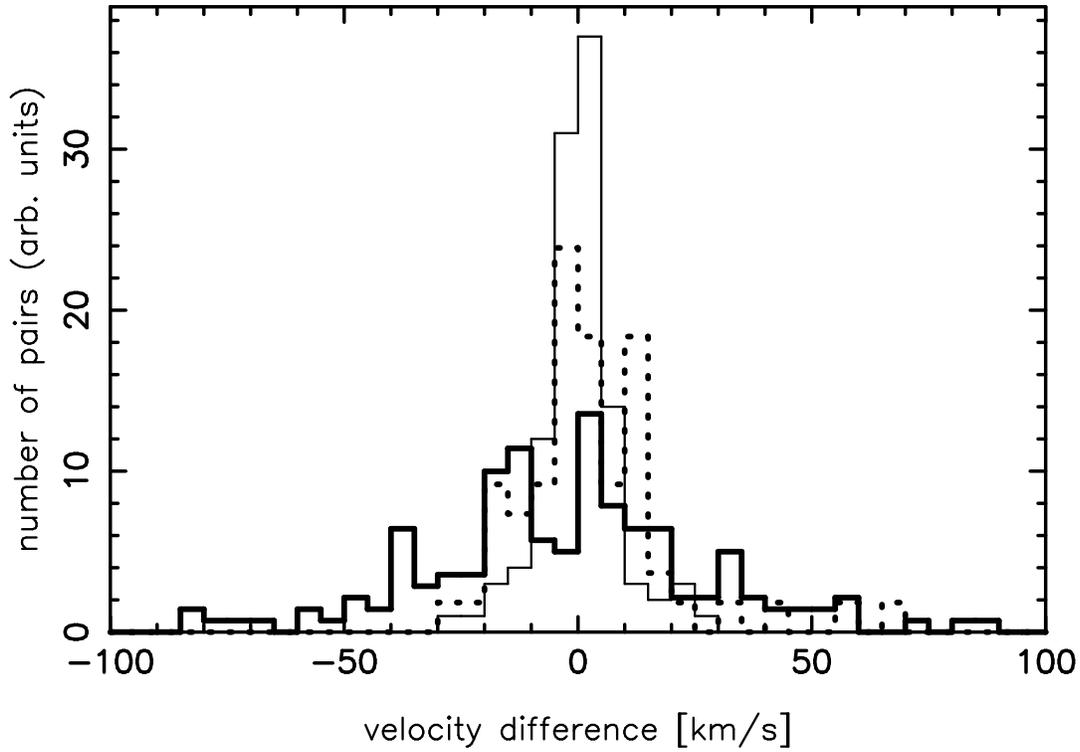}
\caption{\small Observed distributions of velocity shear with the histograms 
from figs.\ref{vdiffs},  \ref{q2345_velewmodels}, and \ref{beckervelewmodels} normalized to the
same integral (arbitrary units) and overplotted on top of each other. The \rxj\ sample is
represented by the thin lined histogram, the \weed\ sample by a dotted one, and the high
redshift combined \patnaik\ and \sdss\ samples by a thick, solid histogram.
\label{overplot3} }
\end{figure}

Fig.\ref{overplot3} shows all three observed shear distributions on the same velocity scale. 
Compared with the Gaussian scatter seen in the case of the very close
lines of sight to \rxj\  
the distributions for the velocity shear in \weed\  at a separation of 60.7 kpc (fig.\ref{q2345_velewmodels})
looks less Gaussian, and the combined distribution for the two higher redshift pairs (fig.\ref{beckervelewmodels})
(\patnaik\ and \sdss) has clearly developed broad wings, not unlike a Lorentzian.
Below we show that this peculiar shape is exactly what is expected for a population
of randomly oriented, highly flattened structures expanding with the general cosmic web.

The distribution histograms as shown in figs. \ref{q2345_velewmodels} and \ref{beckervelewmodels} are incomplete at a level that
depends mainly on confusion as the absorption-line density goes up with
redshift,  and partly on the noisiness of the data.
Confusion happens when two absorption lines in two adjacent lines of sight
are mistakenly ascribed to the same underlying cloud. 
The rate of incidence per unit redshift of absorption lines around redshift 2 is still
small enough that this is not a concern, but beyond redshift 3 the likely
velocity shifts become comparable to the average redshift separation along the line of sight
between absorption lines. Moreover, separations between the lines of sight 
on the order of several hundred kiloparsecs 
as considered here already exceed the typical length over which Ly$\alpha$ absorbers are uniform enough to merit speaking of individual clouds (Cen \& Simcoe 1997).
Then it is difficult to be sure that a given absorption system continues
across the sky in the other line of sight. For the two high-redshift pairs discussed here 
and shown in fig. \ref{beckervelewmodels}, the incompleteness is estimated to set in already at velocity differences of
less than 100 kms$^{-1}$, leading to a systematic underestimate of the width of the
velocity distribution. 
 Below we describe how to correct for these systematic errors
by analyzing simulated Ly$\alpha$ forest spectra from a cosmological simulation in 
exactly the same way as the real data, in an attempt to introduce the same biases and relate
the observed width of the velocity shear distribution to the underlying three-dimensional
kinematics of the
gas.

\subsection{Modelling the Distribution of Velocity Shear as Large Scale Expansion}

To get a qualitative understanding of the observed motions
we first proceed to analytically model the observed shape of the distribution of velocity differences to see whether it is consistent with motions expected of clouds partaking in
the Hubble expansion. Moreover, we 
check whether the order of magnitude of the expansion velocity
is really consistent with this interpretation.

\begin{figure}[p]
\includegraphics*[scale=.6,angle=0.]{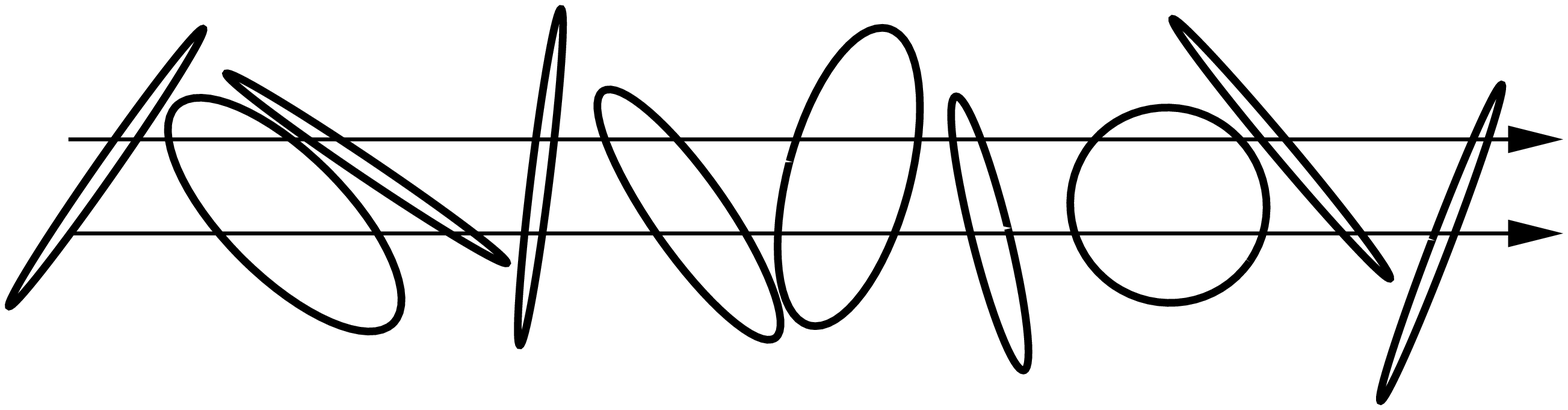}
\caption{\small Randomly orientated, radially expanding pancakes
intersecting two close lines of sight.\label{disks.xfig} }
\end{figure}

In the spirit of Haehnelt (1996) and Charlton et al (1995, 1997) we start with a simple model of the expanding clouds, representing
them as a population of flat circular pancakes, all with the same radius, expanding linearly with varying fractions of the
Hubble flow, and having random inclinations on the sky (fig.\ref{disks.xfig}). 
This model may seem unrealistic (and in fact, it is less sophisticated than 
the similar attempt by Charlton et al.),
but there are several reasons to believe that it is a viable first step toward measuring
the effect we are after, the Hubble expansion of the IGM. First, any sample of absorption
lines is dominated by the objects with the largest geometric cross section, so a pancake
is the best choice for a given radius. 
Second, homologous (i.e., velocity $\propto$ length) Hubble expansion may be a good assumption for sheets in
the general IGM because the overdensities are moderate and structures are not
expected to have collapsed in their longest linear dimension. 
The assumption of only one radius for the
pancakes (as opposed to a distribution of radii) is more questionable, as a finite
radius for a tilted expanding pancake corresponds to an upper limit in the velocity shear
and introduces a cutoff in the distribution of velocity differences, so we need to apply
some caution and not consider velocity shear beyond a certain value.

\begin{figure}[p]
\includegraphics*[scale=1.,angle=0.]{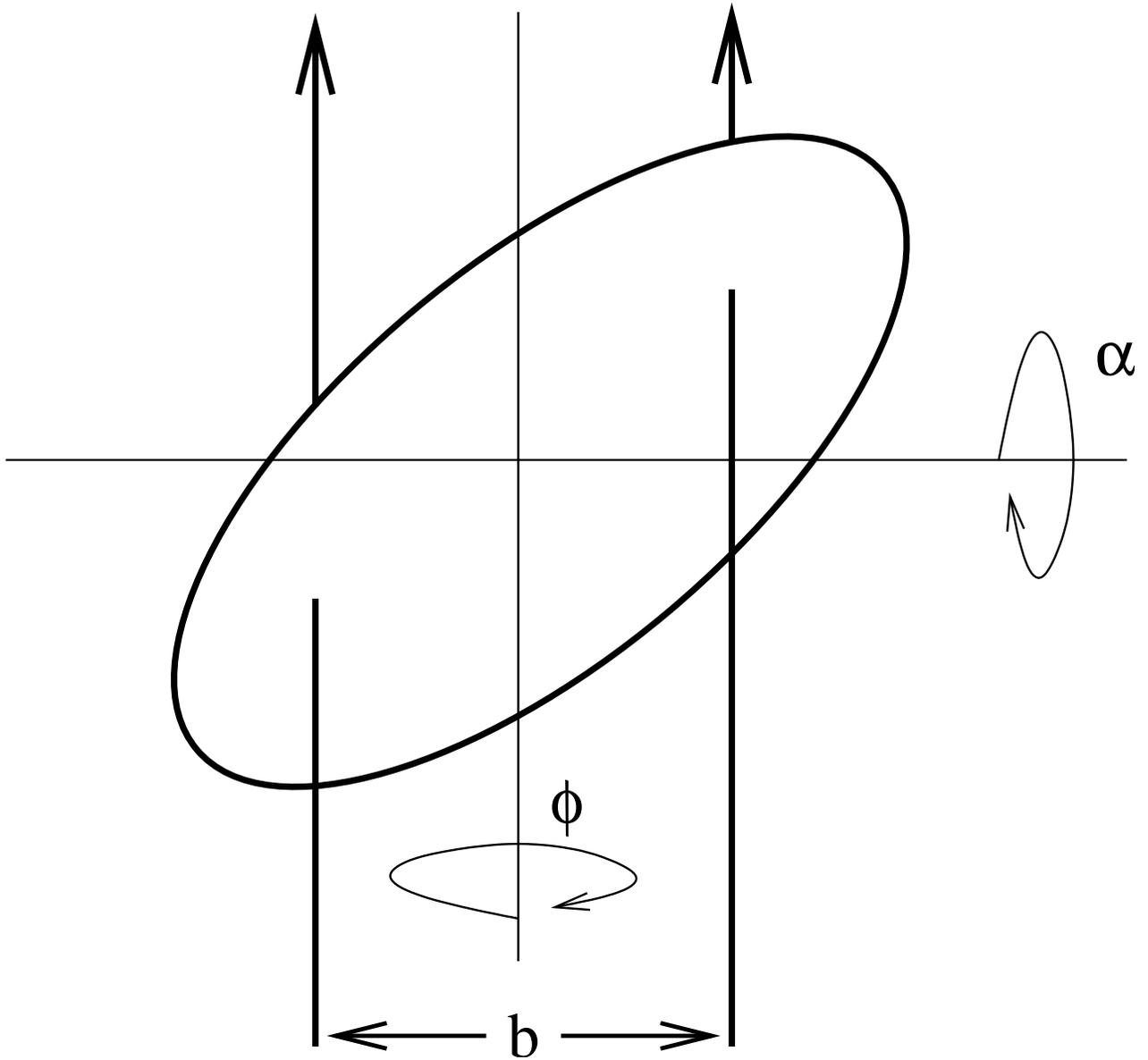}
\caption{\small Homologously expanding pancake intersected by two lines of sight.
The normal vector on the pancake surface is tilted with respect to the direction
of the lines of sight by an angle $\alpha$, and the tilt axis is rotated relative
to the connecting line $b$ between the lines of sight by an angle $\phi$.\label{panc.xfig} }
\end{figure}

The projected velocity
shear $\Delta v$ between two lines of sight separated by a proper beam separation $b$, hitting a circular pancake-shaped cloud that expands radially with expansion velocity $v_{exp} = r H(z) b(z)$ at an inclination angle $\alpha$ ($\alpha$ = 0 would
be face-on) and with an azimuthal angle $\phi$ (fig\ref{panc.xfig}), is given by
\begin{eqnarray}
\Delta v = r H(z) b(z)  \tan{\alpha} \sin \phi. \label{disk}
\end{eqnarray}
Here $H(z)$ is the Hubble constant at redshift z, and the Hubble ratio $r$ is defined as the ratio of the
expansion velocity of the pancakes to the Hubble expansion (i.e., $r=v_{exp}/v_{Hubb}$; $r=1$ would be pure Hubble
flow).

Introducing the angular separation between the beams, $\Delta\theta$, and the angular diameter distance, $D_A$, this can be written
\begin{eqnarray}
\frac{\Delta v}{\Delta\theta} =  r H(z) D_A(z) \tan{\alpha} \sin \phi.
\end{eqnarray}

Adopting the nomenclature used by McDonald \& Miralda-Escud\'e (1998) in their work on the Alcock-Paczynski test,\footnote{There have been a number of suggestions to exploit the Alcock-Paczynski effect 
using paired Ly$\alpha$ forest sightlines to derive cosmological parameters (e.g., 
McDonald \& Miralda-Escud\'e 1998; Hui, Stebbins \& Burles 1999;  Rollinde et al 2003, Lidz et al 2003). Essentially, this measurement employs auto and cross-correlation functions of the absorbed flux in the Ly$\alpha$ forest to measure a function of cosmological parameters (especially $\Lambda$) only.} we split off the cosmological dependence of $\Delta v$ and write it as  
\begin{eqnarray}
f(z) =  \frac{H(z) D_A(z)}{c}. 
\end{eqnarray}
Our relation for the angular velocity shear becomes
\begin{eqnarray}
\frac{\Delta v}{\Delta\theta} =  r c f(z) \tan{\alpha} \sin \phi.
\end{eqnarray}

The idea is now to fit the observed distribution of $\Delta v$ for absorption lines, trying
to reproduce it
with a model population of these pancakes
hit at random orientations by imaginary double lines of sight. The
ratio $r=v_{exp}/v_{Hubb}$ is treated as the free fitting parameter. Note that 
the factor $f(z)$ and thus the width of the distribution of $\Delta v$ 
is independent of the absolute value of the Hubble constant. This is because
the beam separation $b$ is only known to within a factor $h^{-1}$ (the angular
diameter distance that enters in the calculation of $b$ is proportional to $c/H$),
and the velocity shear is proportional to $b H$.
Thus the result of this measurement is the ratio $r$, which tells us about any
departures from the Hubble flow but does not give the value of $H$.
We will assume that $f(z)$ is completely known, i.e., that we know already
the cosmological parameters reasonably well, and we ascribe any departures of $r$
from unity to local departures from the Hubble flow.
In fact, such departures are expected because typical, unsaturated Ly$\alpha$ clouds are moderately overdense and 
are thought to have collapsed in one dimension, and thus should expand anisotropically, on a sufficiently small scale. In general, a  column density limited sample of absorption lines observed across a finite spatial scale smaller that the typical coherence length will never be representative of the free Hubble flow.

\smallskip

A Monte Carlo simulation of pancake-shaped "clouds" was used to create a distribution $P(\Delta v)$ of the velocity shear, given simultaneous hits of the same pancake by both lines of sight. The pancakes' normal vectors were randomly oriented with respect to the sight lines,
and the hits were weighted with the projected geometric cross section subtended by the pancakes.
The  velocity differences projected along the line of sight between the points of the pancake
hit by the lines of sight where gathered to form a theoretical frequency distribution of the
velocity shear.

In practice, equation (\ref{disk}) shows that because of the nature of the Hubble law
there is a degeneracy between line of sight extent and expansion velocity;
a larger velocity of expansion and a smaller tilt of the pancake with respect to the observer
give the same velocity shear as a  smaller velocity of expansion and a larger tilt.
Larger pancakes admit larger tilts leading to a larger $\Delta v$.
The degeneracy is not perfect because of the finite size of the absorbers, but it is clear that
if we wish to extract the velocity of expansion from the observations we need to 
have prior knowledge of the size of the absorbers.  Numerous such measurements
have been done (see section 1). We are using here the compilation by D'Odorico et al (1998), who found the mean proper radius of Ly$\alpha$ clouds to be $R = 412 h^{-1}_{100}$ kpc.
In agreement with earlier work (Crotts \& Fang 1998) these authors found no evidence of redshift evolution in the
mean size. Transforming the D'Odorico et al (1998) values  to the cosmological model used here  gives a mean radius $R = 503.5 h^{-1}_{70}$ kpc,
with $3\sigma$ limits $(407 < R < 628)$ $h^{-1}_{70}$ kpc. 

We model the Ly$\alpha$ forest as homologously expanding pancakes, with a {\it constant} physical radius at all redshifts ($z\sim 2.04 - 3.8$) in our sample.  
The adoption of a constant physical size for an expanding object may sound counterituitive, 
but we are really comparing common absorption systems {\it above a certain column density
threshold} that is given by observational constraints and does not depend on redshift.
Aside from the observational evidence cited above, theoretical  arguments suggest that
the linear, physical  extent $R$ of a Ly$\alpha$ absorber for a given column density depends only weakly
on redshift. The dependence arises mainly through the ionization rate $\Gamma$ according
to $R\propto \Gamma^{-1/3}$ (e.g., Schaye 2001), which does not appear to
change by more than 50\% from redshift 4 to 2 (Rauch et al. 1997b), so that the change
in radius at constant column density  is less than $15\%$.
Thus using a single radius for the model pancakes is not entirely unjustifiable.

\medskip

\begin{figure}[p][p]
\includegraphics*[scale=.6,angle=-90.]{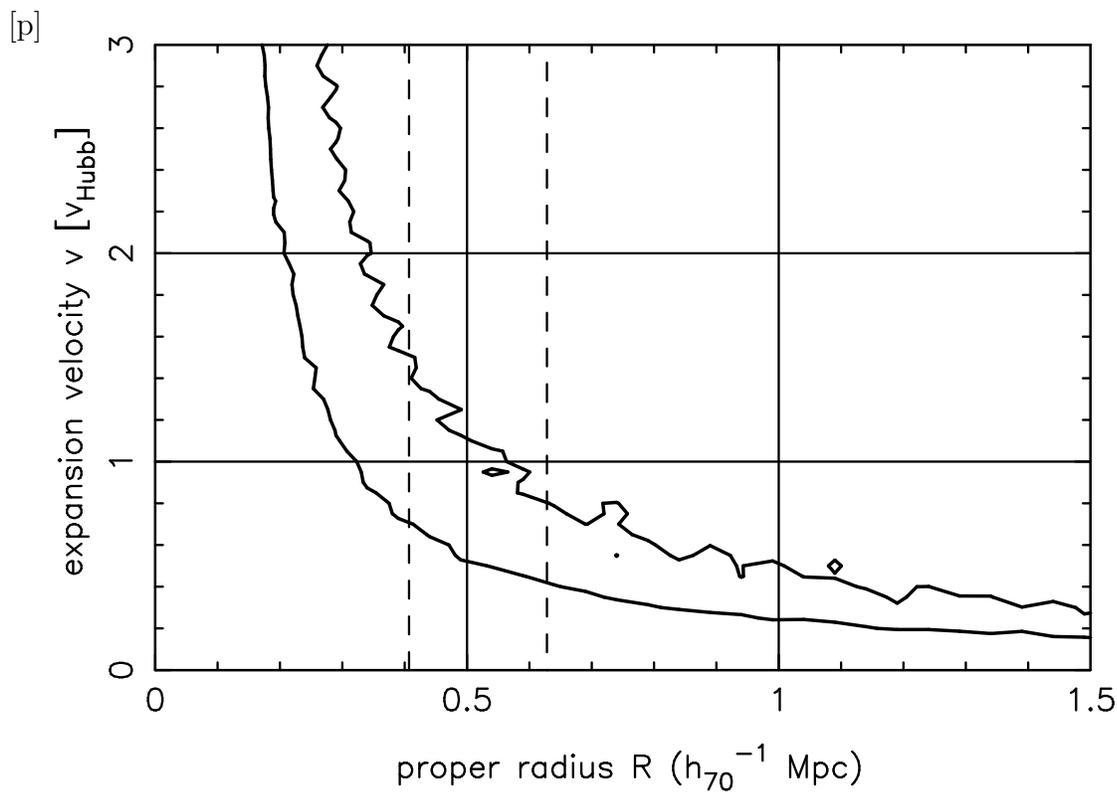}
\caption{\small The solid lines are the $\chi^2$ contours ($\pm 3\sigma$) for the maximum likelihood fit of the expanding pancake model to the velocity
shear distribution at $z\approx 2$ (see fig.\ref{q2345_velewmodels}), with the expansion velocity
in units of the Hubble velocity and the proper radius of the pancakes as free parameters. The vertical dashed lines give the $\pm 3\sigma$ limits for the radii of the absorbing structures from D'Odorico et al (1998). According to this plot, the average expansion
of the Ly$\alpha$ forest at mean redshift $<z> = 2.04548$ is
$v = (0.8\pm 0.3)\times  v_{\mathrm Hubble}$ (approx. $3\sigma$). \label{q2345_chisq_bw}}
\end{figure}

\begin{figure}[p]
\includegraphics*[scale=.6,angle=-90.]{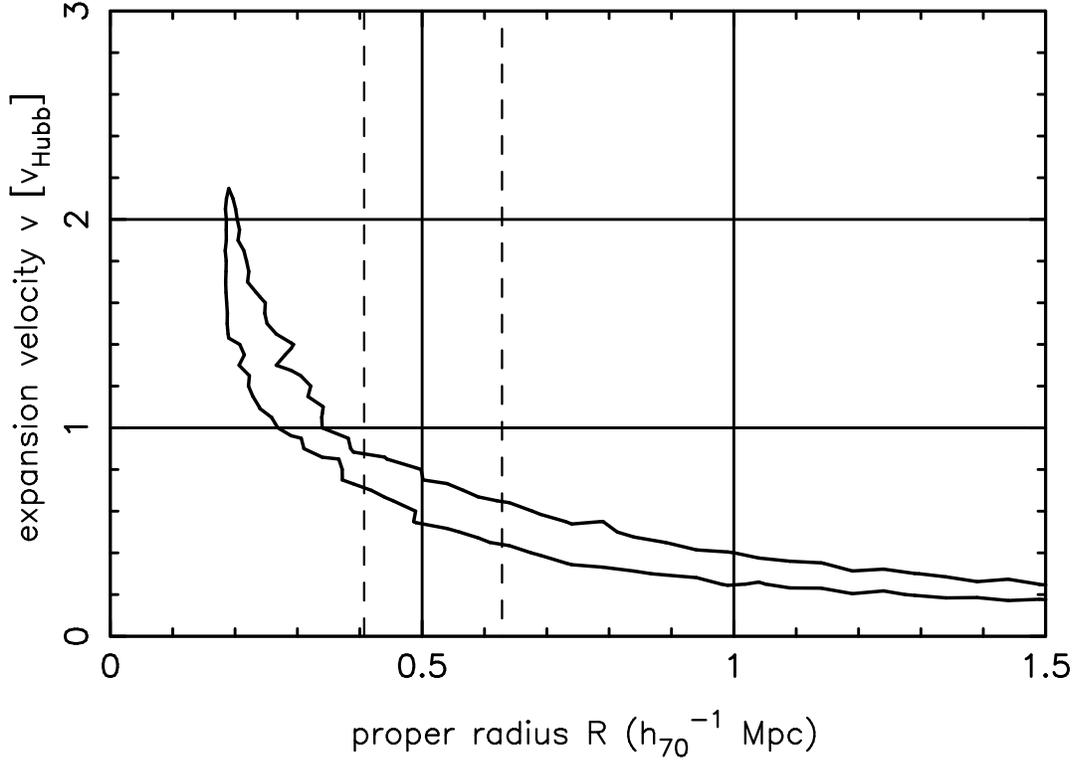}
\caption{\small The solid lines are the $\chi^2$ contours ($\pm 10\sigma$) for the fit of the expanding pancake model to the velocity
shear distribution at mean redshift $<z>=3.53$ (see fig.\ref{beckervelewmodels}). Unlike the
$z=2$ case, in the present case these error contours are meaningless, as the total error is  dominated by the
uncertainty in the cross identification of absorption lines between the lines of sight. The vertical dashed lines again give the $\pm 3\sigma$ limits for the radii of the absorbing structures from D'Odorico et al (1998).  According to this plot, the average expansion
of the Ly$\alpha$ forest at mean redshift $<z> = 3.53$ is
$v = (0.65\pm 0.4)\times  v_{\mathrm Hubble}$. \label{beckerpairs_chisq_bw} }
\end{figure}

The results of maximum-likelihood-fitting the expanding pancake model  to the observed velocity shear
distributions are shown in figs.\ref{q2345_chisq_bw} and \ref{beckerpairs_chisq_bw}. The former gives the $3 \sigma$
$\chi^2$ contours for the best fitting combination of proper radius and expansion velocity
in units of the Hubble velocity for the Q2345+005A,B lines of sight. Adopting the D'Odorico et al (1998) value for the
radius, the best fit for the Hubble ratio is $r= 0.8\pm 0.3$ ($3\sigma$). The corresponding
theoretical curve with that value of $r$ is shown overplotted as a solid line in fig.\ref{q2345_velewmodels}.
For comparison, the curves for $r=0.4$ ({\it dashed line}) and $r=1.5$ ({\it dotted line}) are also 
shown. The $r=0.4$ value produces a distribution too centrally condensed, whereas the
higher value $r=1.5$ gives too strong wings for the distribution. The main uncertainty in
this (redshift $\sim 2$) case comes from the finite number of absorption systems  available in the spectrum.
The fit for the higher redshift samples is given in fig. \ref{beckervelewmodels}, with a formal
best-fit value of $r=0.65$. Here the statistical
errors are small (we are showing the $10 \sigma$ contours!) but the main (and systematic)
uncertainty comes from the confusion between unrelated systems and from missing the largest
velocity separations. These effects are not taken into account  in producing the $\chi^2$ contours. From looking at individual absorption-line systems and redoing the 
line selection repeatedly on different subsets of the data, we estimate that the value
could well be between 0.4 and 1.2, and we show these curves overplotted on the observed histogram in fig.\ref{beckervelewmodels}, but even this error estimate itself is uncertain.

A better assessement of the reliability of these estimates of the Hubble ratio requires
a more realistic model for the IGM, which we provide in the following section.
We note, however, 
that a very simple model such as the expanding pancake reproduces
the basic shape of the observed distribution of velocity shear quite well, and in combination with
the best estimate of the coherence length for the Ly$\alpha$ forest clouds it gives
values of the expansion velocities for the moderately overdense IGM  relatively
close to the Hubble expansion.

\section{A Comparison with Cosmological Hydro-simulations}

To be able to gauge the meaning of our measurements of velocity shear
(figs. \ref{q2345_velewmodels} and \ref{beckervelewmodels}), and to understand how the velocities arise, we produced artifical Ly$\alpha$ forest spectra from a numerical
cosmological hydrodynamic simulation. 
In such a simulation, the density and velocity field are of course known per definition,
and it becomes possible to invert (in a primitive sense, at  least) the spectrum to
see which combinations of density, peculiar velocity, and Hubble expansion  
conspire to form a given absorption line, and, in close lines of sight, a pair of those. In particular, one can ask the questions,
How do to the physical structures (gaseous filaments, etc.) expand or contract in order to give a
certain distribution of velocity shear ? And how do the considerable selection effects in the
spectral domain
affect the measurement of their velocities ?

We use a new version of the parallel tree SPH (smoothed particle hydrodynamics) code GADGET
(Springel et al 2001) in its tree PM (particle mesh) mode, which speeds up the
calculation of long-range gravitational forces considerably. The
simulation is performed with periodic boundary conditions with
$400^3$ dark matter and $400^3$ gas particles. Radiative cooling and
heating processes are followed using an implementation similar to
Katz et al (1996) for a primordial mix of hydrogen and helium. The UV
background is given by Haardt \& Madau (1996). In order to maximize the speed of
the simulation, a simplified criterion of star formation has been
applied: all the gas at overdensities larger than 1000 times the mean
overdensity is turned into stars (Viel et al 2004).
The simulation was run on {\sc cosmos}, a 152 GB shared memory Altix
3700 with 152 CPUs hosted at the Department of Applied Mathematics and
Theoretical Physics (Cambridge).
The cosmological parameters are
$\Omega_{{\rm M}}= 0.26$, $\Omega_{\Lambda} = 0.74$, $\Omega_{{\rm B}} =
0.04
63$ and $H_0=72\,{\rm km\,s^{-1}Mpc^{-1}}$.  The $\Lambda$CDM transfer
functions have been computed with
{\sc cmbfast} (Seljak \& Zaldarriaga 1996).

The comoving size of the box was 60 $h^{-1}$ Mpc.
At three different redshifts (z=2, 3.4, and 3.8, close to the mean redshifts
in the observations), 20  artifical lines of sight of
lengths 5571, 6533, and 6789 kms$^{-1}$ were run through the simulated box.
The effective HI optical depth of the spectra was adjusted so as to match
the phenomenological fitting formula given by Schaye et al (2003) for each redshift.
The lines of sight were created in pairs with transverse separations identical
to the mean separations in our three observed QSO pairs, and there were 10 fake "QSO pairs"
at each redshift.  
The Ly$\alpha$ forest spectra were subjected to the same analysis as the real
data; i.e., spectral regions with assumed common absorption features in each
pair were selected by eye and delineated with a cursor. A uniform minimum rest-frame
equivalent width threshold of 0.4 \AA\ was imposed, and the flux-weighted line-of-sight velocities were calculated. 

Then all the spatial pixels along the line of sight whose total (= peculiar + Hubble) velocity projected
into one of the selected absorption-line windows were identified. Their spatial
positions (weighted by the square of the gas density, to emulate their contribution 
to the absorption-line optical depth) were used to obtain the spatial "centroid" along the 
line of sight of the 
gas clump causing the absorption in each pair spectrum. This procedure is crude in three ways:
it ignores thermal motions and small-scale turbulence; it takes the recombination
rate ("square of the density") as a proxy for optical depth; and it assumes that the
Ly$\alpha$ forest lines typically are caused by overdensities, as opposed
to velocity caustics (e.g., McGill 1990). The two former simplifications are clearly justified 
by us only attempting to measure the global shifts between entire absorption lines.
The identification of most absorbers with overdensities (and rarely velocity caustics) is consistent with results
from previous simulations (e.g., Miralda-Escud\'e et al 1996).

\begin{figure}[p]
\includegraphics*[scale=.6,angle=0.]{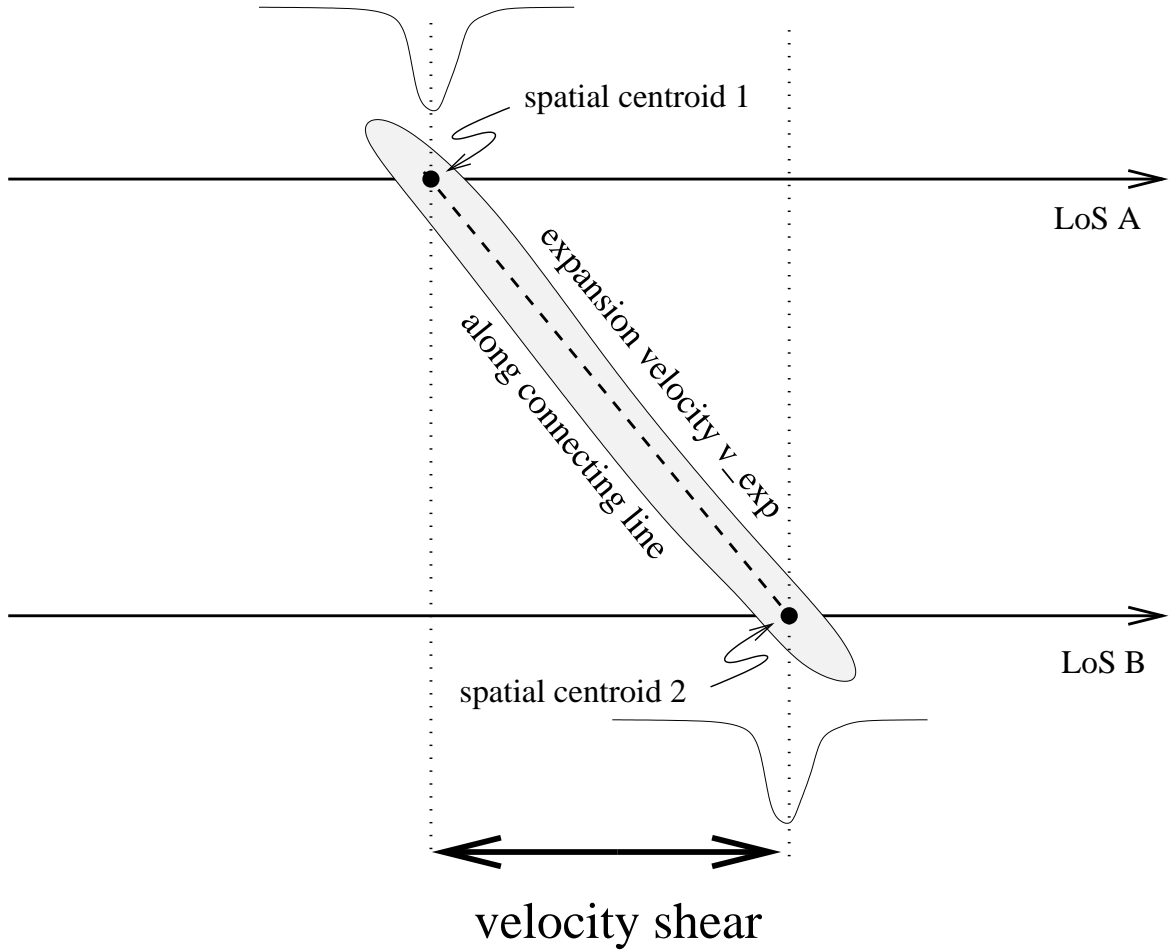}
\caption{\small  Illustration of the method for relating the velocity shear along
the lines of sight to the expansion velocity of an absorption-selected gas cloud.
(The figure is strictly valid only for pure Hubble flow where the Hubble law guarantees
that the angles and positions are the same in position space and velocity space).
Figs. \ref{q2345_velewmodels}, \ref{beckervelewmodels}, \ref{sim_dat_2345}, and 
\ref{sim_dat_1422_1439} show the velocity shear, whereas figs. \ref{z_3.8_hubrat},
\ref{z_3.4_hubrat}, and \ref{z_2_hubrat} show the distribution of the expansion
velocity along the connecting line between the spatial centroids.
\label{crosslos.xfig} }
\end{figure}

Having obtained the spatial centroid where the density clump contributing most to a given absorption line intersects  the two lines of sight, the relative three-dimensional velocity vector between these two positions
is computed from the Hubble expansion and peculiar velocity array (fig.\ref{crosslos.xfig}).
Thus, for each common absorption system in a pair of lines of sight, we know
the three-dimensional relative velocity between the parts of the absorbing structure intersecting
the lines of sight. It becomes possible to relate the observed, one-dimensional distribution
of velocity shear to the three-dimensional motions of the IGM.

\medskip

\begin{figure}[p]
\includegraphics*[scale=.6,angle=-90.]{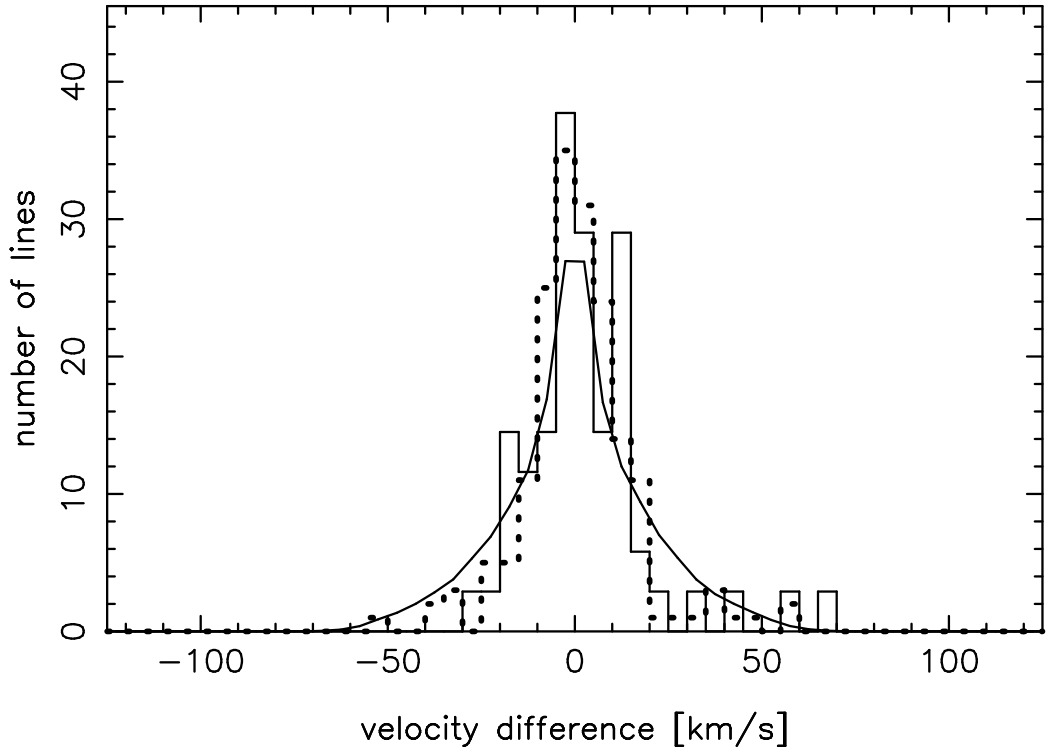}
\caption{\small  Observed distribution for Q2345+005 of $\Delta v$ (solid histogram), distribution
from hydro-simulation for z=2 (dotted histogram) and the best fitting expanding pancake model, expanding with 0.8 times
the Hubble velocity.\label{sim_dat_2345} }
\end{figure}

\begin{figure}[p]
\includegraphics*[scale=.6,angle=-90.]{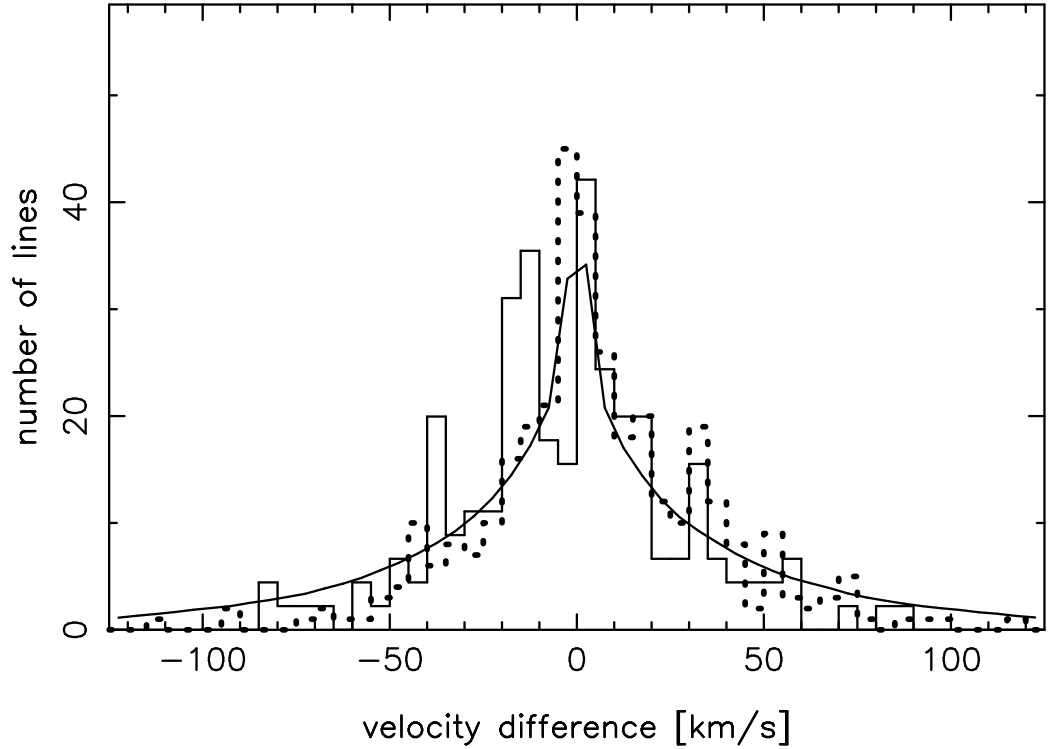}
\caption{\small Same as Fig. 19, but with the solid histogram now showing
the $\Delta v$ measurements for the higher redshift ($<z>\approx 3.6$) combined
sample from  the
Q1439-0034A and B pair,
and 
the Q1422+2309A and Q1422+2255 pair. Again the dotted line is from the simulation
for mean redshift $<z>=3.6$, and the solid curve is the expanding pancake model for the same redshift
with a Hubble ratio $r= 1.0$.\label{sim_dat_1422_1439}}
\end{figure}

The resulting {\it simulated} shear distributions for $z=2$ and $z=3.6$ (the samples for $z=3.4$ and 
$z=3.8$ were combined to increase the statistics) are plotted as dotted histograms on top of the {\it real} data
(same as in figs.\ref{q2345_velewmodels} and \ref{beckervelewmodels})
in figures \ref{sim_dat_2345} and \ref{sim_dat_1422_1439}. The only adjustment applied was for the integral of the curves to be the same. 
A Kolmogorov-Smirnov (K-S) test shows that the observed and simulated unbinned cumulative
distributions of velocity shear are consistent with each other in the usual sense; i.e., the
maximum differences between the cumulative distributions are expected to be exceeded in 30\% 
($z=2$) and 15\% ($z=3.6$) of all random realizations, respectively. 
The rms velocity widths of the distributions are 16.6 kms$^{-1}$ (observed) versus
14.9 kms$^{-1}$ (simulated) in the redshift $z=2$ case, and 30.0 kms$^{-1}$ (observed) versus
30.6 kms$^{-1}$ (simulated) in the redshift $z=3.6$ case. The results are summarized in
table 1.
The sample sizes  are unfortunately not very impressive, 
but they are clearly enough to rule out underlying differences between the widths of the observed and 
simulated distributions at the 50\% level.
We conclude that the simulation reproduces both the observed {\it average} velocity shear and the observed {\it shape}
of the one-dimensional distribution in the IGM quite well.

\subsection{The Theoretical Distribution of Expansion Velocities}

\begin{figure}[p]
\includegraphics*[scale=.6,angle=-90.]{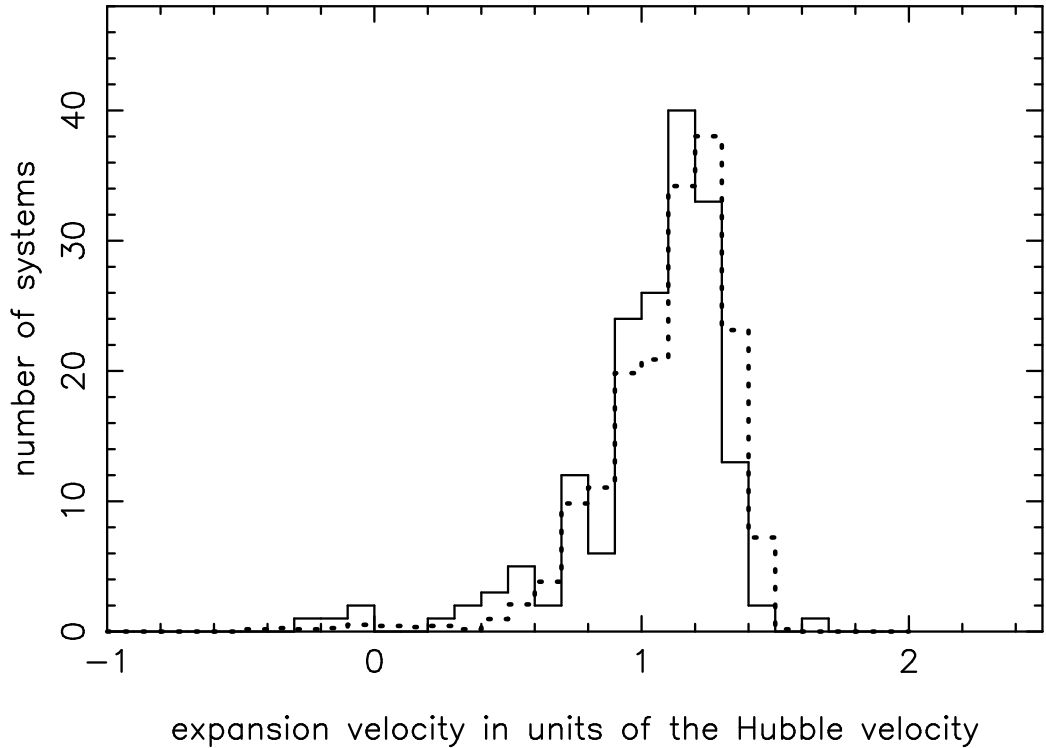}
\caption{\small Expansion velocity in units of the Hubble velocity, along the straight line
connecting the spatial absorption centroids (see fig.\ref{crosslos.xfig}), 
in the z=3.8 simulation (histogram with solid lines). Note the "super-Hubble" peak and the "sub-Hubble" tail
indicating break-away from the general expansion.  Most clouds expand somewhat
faster than the Hubble flow, but some have broken away and are even contracting.
The dotted histogram applies to the expansion velocity measured at random positions 
along the line of sight (i.e., irrespective of there being an absorption line).
There are only small differences between absorption-selected and random distribution (see text).  \label{z_3.8_hubrat}}
\end{figure}

\begin{figure}[p]
\includegraphics*[scale=.6,angle=-90.]{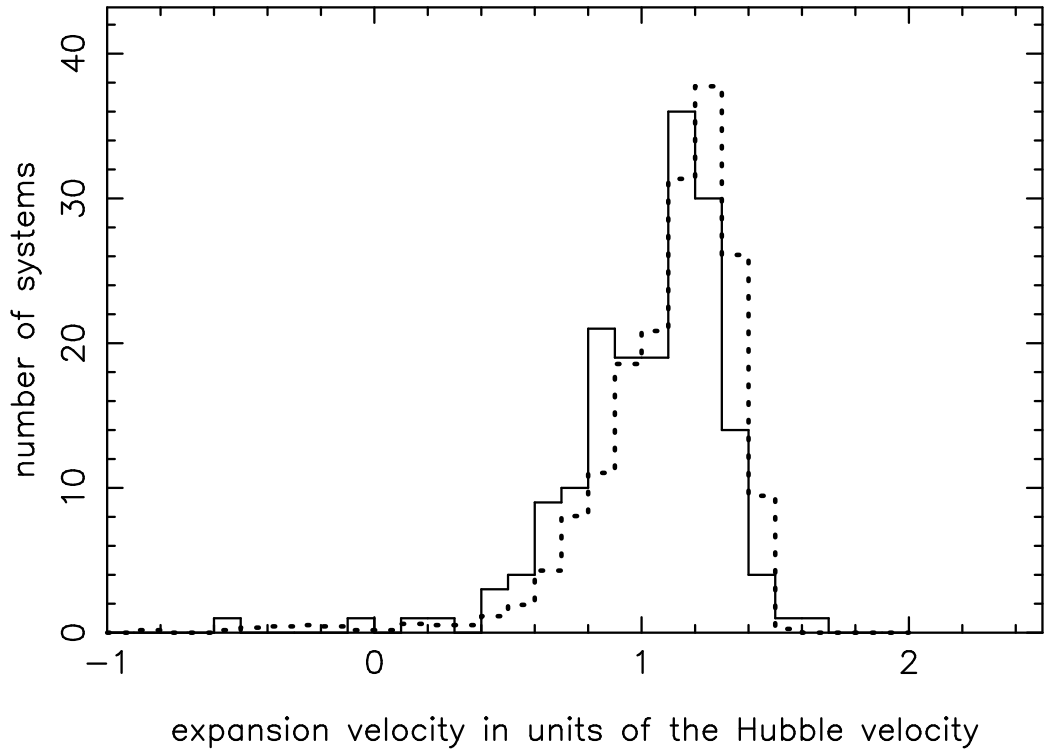}
\caption{\small Same as Fig. 21, but for redshift 3.4. There is little
change since redshift 3.8. However, gravitational collapse here has more noticeably decelerated
the absorption-selected regions relative to the random ones. Note that there is a difference
in redshift {\it and} beam separation between this plot and the previous and following ones.
\label{z_3.4_hubrat}}
\end{figure}

\begin{figure}[p]
\includegraphics*[scale=.6,angle=-90.]{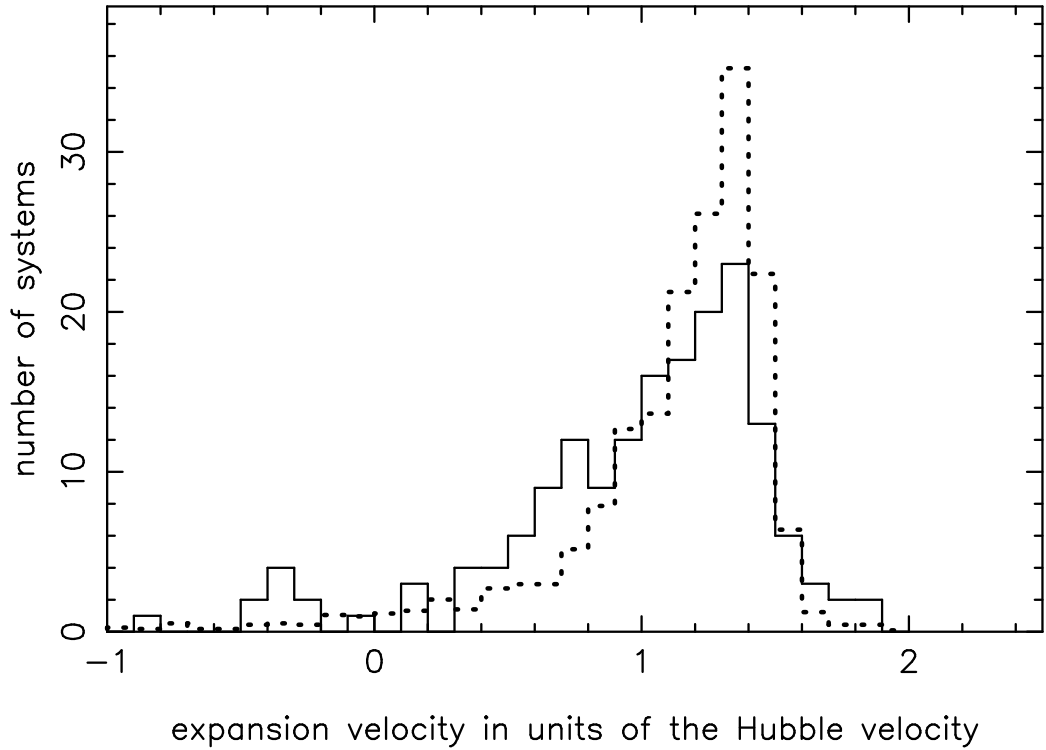}
\caption{\small By redshift 2 and at the smaller (61 kpc) separation, gravitational collapse has broadened the distribution of Hubble ratios in absorption-selected regions (solid histogram) and there are now many regions expanding
faster or slower than the Hubble flow. Random regions (dotted histrogram) are
more dominated by super-Hubble velocities characteristic of voids.
\label{z_2_hubrat}}
\end{figure}

How does the underlying three-dimensional distribution of expansion velocities in the simulation look ? Figures  \ref{z_3.8_hubrat}, \ref{z_3.4_hubrat}, and \ref{z_2_hubrat} give the distributions of the simulated expansion velocities
for redshifts 3.8, 3.4, and 2.0, respectively. To reiterate, these are the relative velocities of  the two spatial
centroids (along the line of sight) of  gas clouds intersected by both lines of sight.

All three diagrams have some features
in common. First, the most probable expansion velocity is larger than the Hubble expansion. The peak of the distribution falls into the $r_{peak}$ = 1.15 (1.15, 1.35) bins for the
three redshifts. The median Hubble ratio is also larger than unity ($r_{med}$ = 1.11 (1.09, 1.08)).  There is a tail toward lower expansion velocities, even including a few physically contracting
systems (with  negative velocities).  Interestingly, the tail grows more substantial with
decreasing redshift, with the mean Hubble ratio going from $r_{mean}$ = 1.03 to 1.02 to 0.85 by
redshift 2. This explains why the width of the $z=2$ observed distribution of {\it velocity shear} seemed narrower
than expected for pure Hubble expansion and why our $3 \sigma$ estimate of $r=0.80\pm0.3$
from the expanding pancake model was smaller than unity (realistically, as it turns out).

For the higher redshift ($z\sim 3.6$), larger separation sample, 
the pancake model seems to have problems, though.
As noted above, the  mean Hubble ratio in the simulations (which give a velocity shear distribution very similar to the one from the real data) is above unity,
but the fit with the pancake model at that redshift gave only an underestimate of $r=0.65$. 
Most likely, the assumption of a non-evolving size
for the pancakes, the finite sizes (relative to the transverse separations between the
lines of sight), and confusion when cross-identifying absorbers, and thus incomplete
counts at the largest velocity differences, are to blame.

\medskip

In any case, the good agreement between the observed
and theoretical distributions (figs.\ref{sim_dat_2345} and \ref{sim_dat_1422_1439}) indicates that the same astrophysical mechanisms at work in the simulation are also present in nature; we are
seeing direct evidence for break-away from the Hubble  flow
and for gravitational collapse in some systems, the number of which increases dramatically
when going to lower redshift.

In contrast to that, most Ly$\alpha$ forest systems continue to undergo super-Hubble expansion
at all redshifts considered here. Filamentary or pancake-shaped structures expanding with
super-Hubble velocities are a natural prediction of CDM-dominated structure formation scenarios (e.g., Haehnelt 1996) and have been proposed to be responsible
for some of the largest velocity structures seen occasionally among metal absorption systems (Rauch et al 1997a). These filaments  occur at the 
boundaries of underdense "voids" that themselves expand faster than the Hubble flow. Another way of
explaining super-Hubble expansion recognizes that
filaments are being gravitationally stretched by and draining into the high-mass nodes terminating them
(presumably future galaxy clusters), thus introducing super-Hubble velocity gradients.

We caution that the numerical results and the distributions given here are obtained in a highly
selective way: admitting only Ly$\alpha$ clouds with rest-frame equivalent widths
above 0.4\AA\ selects denser gas at lower redshift that may be in a more advanced stage of collapse.
In addition, the measurements differ simultaneously in redshift {\it and} beam separation (with mean physical separations of 236, 288, and 61 $h_{72}^{-1}$kpc for $z = $ 3.8, 3.4, and 2.0). The expansion velocities are measured along straight lines
between the density centroids selected by the absorption systems that they cause, so they do not take into
account any curvature of the clouds, especially at the larger separations.
Therefore the three histograms may be representing different density regimes, size scales, and cloud
shapes at the three
redshifts. They {\it do not necessarily correspond to an evolutionary sequence}.

\medskip 

We defer an assessment of the various selection effects and a discussion of
the physical properties of the absorbers in the simulation to a future paper, but we
can briefly ask the following question: in what sense do the motions of the objects in the simulation selected
by their Ly$\alpha$ forest absorption differ from those of random regions in the universe ?
To construct a control sample of "random regions" we calculated the Hubble ratios 
between random (i.e., not absorption-selected) points along one line of sight and
corresponding points in the "partner" line of sight at directions from the former
that were drawn randomly from the distribution of orientations between the absorption-selected
points. The results are overplotted as dotted histograms in figs.\ref{z_3.8_hubrat},  \ref{z_3.4_hubrat}, and \ref{z_2_hubrat} and summarized in table 2.

There is little difference at redshift 3.8, but
already by $z=3.4$ and much more so by $z=2$ the distributions of the Hubble ratios
have shifted considerably between random and absorption-selected regions. The mean Hubble ratios at redshifts 3.8, 3.4, and 2
are 1.03, 1.02, and 0.85 (absorption-selected) and 1.09, 1.09, and 1.08 (random),
whereas the median Hubble ratios were 1.11, 1.09, and 1.08 (absorption-selected)  and 1.15, 1.16, and 1.22 (random).  Obviously, the overdense regions selected by the Ly$\alpha$ absorption
are undergoing gravitational collapse faster than the random places. This is not
suprising, as a random position in the universe is most likely to end up in underdense
regions that expand faster than the Hubble flow. Note that the median Hubble ratio for
the random regions is even increasing with decreasing redshift, presumably because it
becomes harder to hit overdense regions with an ever-decreasing cross section.

\section{Limits on Other Sources of Motion in the IGM}

Aside from pure Hubble expansion and motion in a gravitational
potential well, one may expect galactic feedback, including
galactic outflows, thermal expansion, or radiation pressure, or other hydrodynamic effects like
ram pressure stripping,  to contribute to the motions in the IGM. 
There is now clear evidence that some of the above feedback processes must have led to
widespread and early metal enrichment in the IGM. By redshift 3, much of the Ly$\alpha$ forest is metal-enriched  (e.g., Cowie et al 1995, Tytler et al 1995, Ellison et al 2000; Schaye et al 2000,2003; Songaila 2001; Simcoe et al 2004). There is also evidence, at least for the
stronger metal absorption systems, of recent injection of turbulent energy in the IGM, at the
level of both the individual absorption lines and the entire absorption
complexes (Rauch et al 1996, 2001a).
These findings point to the importance of the interactions  between galactic potential wells and their IGM environment.

Of the above effects, galactic superwinds have perhaps received the most attention.
These winds have primarily been seen  close to the starforming regions they
originate in (McCarthy et al 1987; Franx et al 1997; Pettini et al 2001, 2002; Heckman
 2002), but   based on their large power and analogies with
low-redshift superwinds it has been proposed  that they may be able 
to escape galaxies and profoundly affect the properties of the IGM, blowing bubbles of highly ionized, metal-rich
gas out to distances of more than half a Mpc (comoving;
Adelberger et al 2003; Cen et al 2005).
The first instances of individual superwinds actually {\it leaving} high redshift galaxies 
may have been seen in MgII (Bond et al 2001a,b) and OVI (Simcoe et al 2002) absorption systems. It is less clear whether such winds would be common and/or strong enough to significantly
alter the properties of the IGM. Simulations suggest that their impact may mostly affect very high column density
systems (with neutral hydrogen column densities $N_{HI} > 10^{16}$cm$^{-2}$; Theuns et al 2001).
Searches for proposed signatures of cosmological
wind shells (Chernomordik 1988) in the autocorrelation function of the Ly$\alpha$ forest
have not been successful (Rauch et al 1992). Employing differential measurements across
close lines of sight, Rauch et al (2001b) concluded that the general IGM (unlike rare, strong
metal absorption systems) does not show the small scale density structure likely to be associated
with the recent passage of winds across the lines of sight.

It is tempting to revisit this question here and see whether the {\it velocities} in the
IGM can shed new light on the impact (or otherwise) of such superwinds. 
We first briefly consider the likely
observational signature of such outflows in the general IGM, and 
then ask specifically the question as to whether the observed velocity
distribution of Ly$\alpha$ forest clouds can be affected by  winds.

\subsection{The Observability of Cosmic Superwinds}

While the actual wind material from superwinds is too hot to
be seen in absorption by UV resonance lines, there are a number
of ways in which winds may be associated with lower ionization gas detectable
as QSO absorption lines: winds may produce shells of swept-up IGM gas;
they may entrain and expel ISM; they may disturb the density field
of the ambient IGM and impart kinetic energy to it.

The detection of winds in the Ly$\alpha$ forest employed here relies on measuring these effects as differences between the appearance
of the same absorber between multiple, relatively close sight lines.

Rauch et al 2001b used "cosmic seismometry" (i.e., expected transient 
differences in optical depth or column density between adjacent lines
of sight ) to limit the filling factor of winds in a simple toy model.
Column density variations across the lines of sight may conceivably arise either
directly from the passage of wind material, when small-scale entrained matter or a swept-up shell of IGM are intersected, or they may appear when the undisturbed external
IGM gets hit by the shock front. For the range of mechanical energy and
the ambient density associated with typical galactic superwinds, the swept-up shell
should in principle contain enough HI to be seen in absorption (if it is photoionized), but the
detectability  depends crucially on the
ionization mechanism. The cooling times for very energetic winds may be too
long to produce a lot of neutral hydrogen (e.g., Bertone et al 2005), and one may have to resort to observing higher ions
(e.g., OVI; Simcoe et al 2002).
The entrained matter should, however, be more easily visible in absorption
because of the high density of the ISM where it came from, and because it is 
likely to dominate the mass of the ejecta (e.g., Suchkov et al 1996). 
It is not clear whether results from low-redshift
winds provide any reliable guidance  to $z\sim 3$ winds, but such observations show that the entrained material is even visible in the NaI $\lambda\lambda$ 5890, 5896 \AA\ doublet
(e.g., Phillips 1993; Rupke et al 2002; Martin 2005). Depending for its formation
on largely neutral gas, NaI is one of the rarest ions seen in QSO absorption spectra.
If it is present, many other ions are likely to be much more conspicuous.

\subsection{The Kinematic Signature of Superwinds}

\begin{figure}[p]
\includegraphics*[scale=1.,angle=-0.]{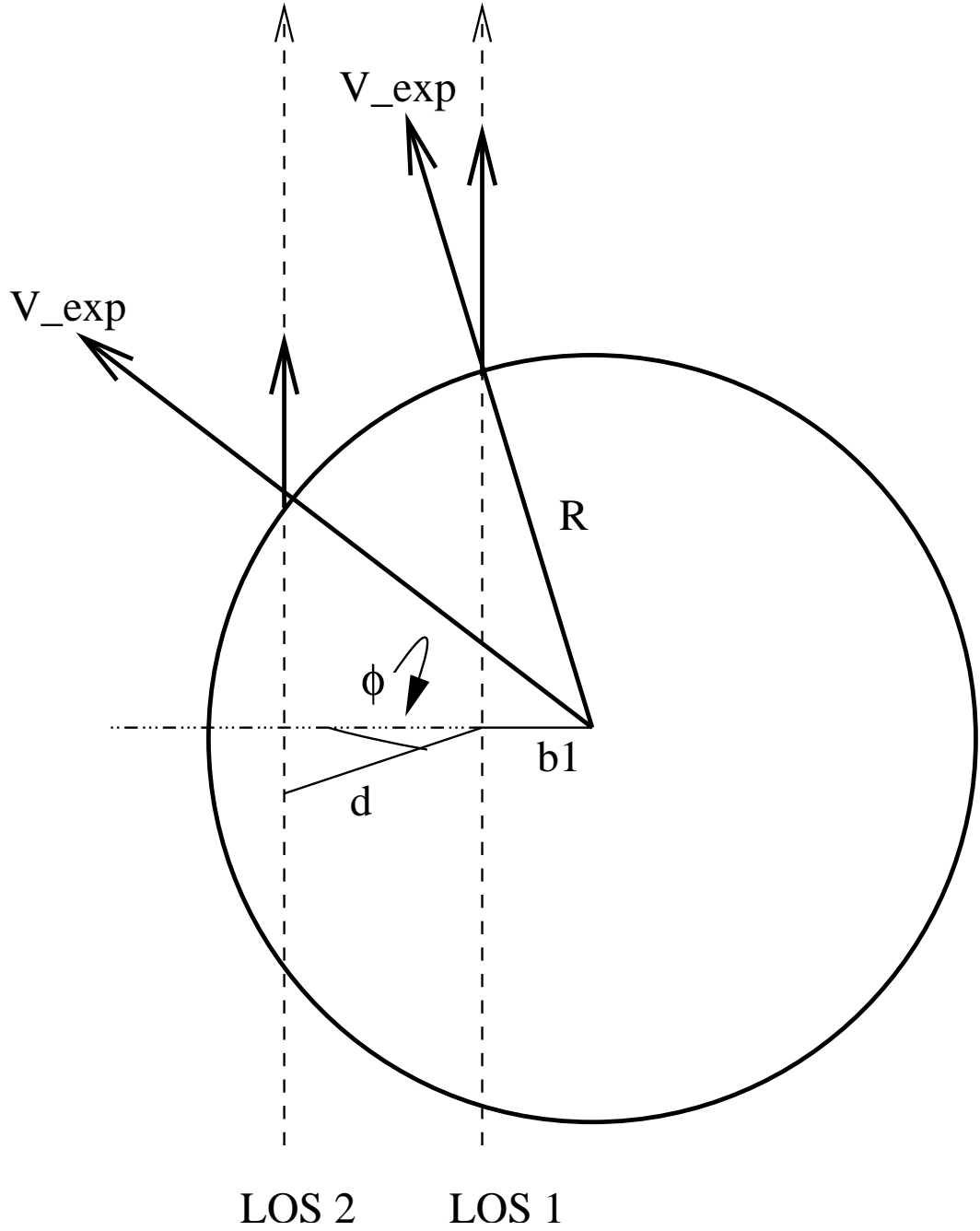}
\caption{\small velocity shear arising from projection effects when an expanding bubble is intersected by two lines of sight
to background QSOs \label{expand.xfig}}
\end{figure}

Even if most superwinds were simply materially
invisible in absorption and would not produce cold shells, or would evaporate all the
entrained matter, 
the expanding bubble should have a kinematic impact on the surrounding IGM and accelerate the ambient HI containing gas.  The acceleration should lead to detectable shifts
of the absorption lines caused by clouds in the path of the wind,
independently of whether they were produced by the wind (as cooling shells or entrained gas) or were present already
before, e.g.,  in the form of gravitationally collapsed filaments. To estimate the order of magnitude of the wind velocities consistent with the observations,
we adopt a simple model where  a spherical shell of gas is pushed radially outward 
by a wind. It is pierced by two randomly oriented
lines of sight and shows up observationally 
in the form of two absorption lines shifted relative to each other
in each of the two  lines of sight (fig.\ref{expand.xfig}). We focus our attention on the
comparison of velocity differences arising on the same side of the shell, as a wind bubble wall
is more likely to be spatially coherent over small distances than at opposing sides of a bubble.
However,
in the case of a spherical bubble the optimal transverse beam separation that maximizes
the observable velocity shear is of course on the order of the
radius of curvature of the wind front, which, for galactic superwinds
may be tens of kiloparsecs. Beam separations much smaller than that would show only small
velocity differences.

Thus, with a separation on the order of 60 physical kpc between the
lines of sight,  \weed\ is the most suitable QSO pair in our sample because it is comparable to the radii of shells proposed
to exist around Lyman break galaxies and should
deliver the strongest constraints on the presence of velocity shear. 

\medskip

The observed distribution of velocity  differences between
the absorption systems in the lines of sight to  \weed\  was given in fig.\ref{q2345_velewmodels}. The observed rms velocity differences for \weed\ are
$\Delta v$ = 16.6 kms$^{-1}$, and this number can serve as the upper limit on the admissible velocity shear
from winds. As discussed above, we assume that these velocity shifts
are caused by Ly$\alpha$ forest absorbers pushed around by 
winds.  
We model the absorbers as spherically expanding, gaseous shells with radius  $R$ and 
expansion velocity $v_{exp}$. 
The expected velocity shear
$\Delta v$ between the projected velocities of absorption lines measured between two lines of sight intersecting 
a shell can be written as
a function of $R$, $v_{exp}$  and various
geometric quantities, \begin{eqnarray}
\Delta v = v_{exp} \left(\sqrt{1-\frac{b_1^2}{R^2}} - \sqrt{1 - \frac{b_1^2 + d^2 - 2 b_1 d \cos{\phi}}{R^2}}\right).
\end{eqnarray}

Here $b_1$ is the impact
parameter of one of the lines of sight with respect to the center of the shell, $d$
is the transverse separation between the lines of sight, and $\phi$ is the angle giving
the relative orientation of the lines of sight with respect to the line connecting
the first line of sight to the center of the shell. 

\begin{figure}[p]
\includegraphics*[scale=.6,angle=-90.]{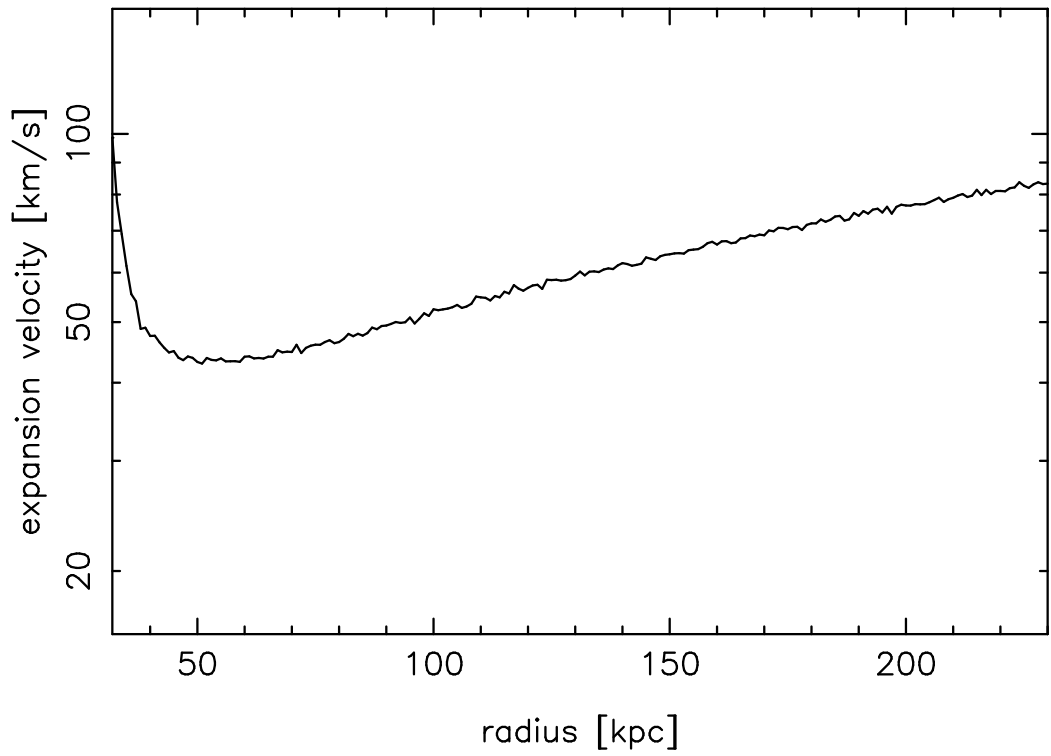}
\caption{\small Relation between radius and expansion velocity for expanding
$z=2$ bubbles capable of producing the mean of the observed distribution of velocity differences,
16.6 kms$^{-1}$.\label{avvexpr}}
\end{figure}

First we ask which combination of radius and expansion velocity gives the same rms velocity difference as the observations. 
We have calculated the quantity $\Delta v$ for a range of bubble radii
from 30 to 230 kpc proper, based on a Monte Carlo simulation
of impact parameters and relative orientations between the 
two lines of sight.  The results are given in fig.\ref{avvexpr}.
Any wind bubble with a radius-velocity combination on this graph
will give a distribution of velocity differences with rms = 16.6 kms$^{-1}$,
as observed.
The admissible expansion velocities range between about 45 and 85 kms$^{-1}$, and have to be compared to the $v_{exp} \sim 600$ kms$^{-1}$ and radii of up to 125 kpc (proper) proposed for winds strong enough to deplete the neutral hydrogen
around Lyman break galaxies by evacuating the HI gas (Adelberger et al 2003). 

\begin{figure}[p]
\includegraphics*[scale=.6,angle=-90.]{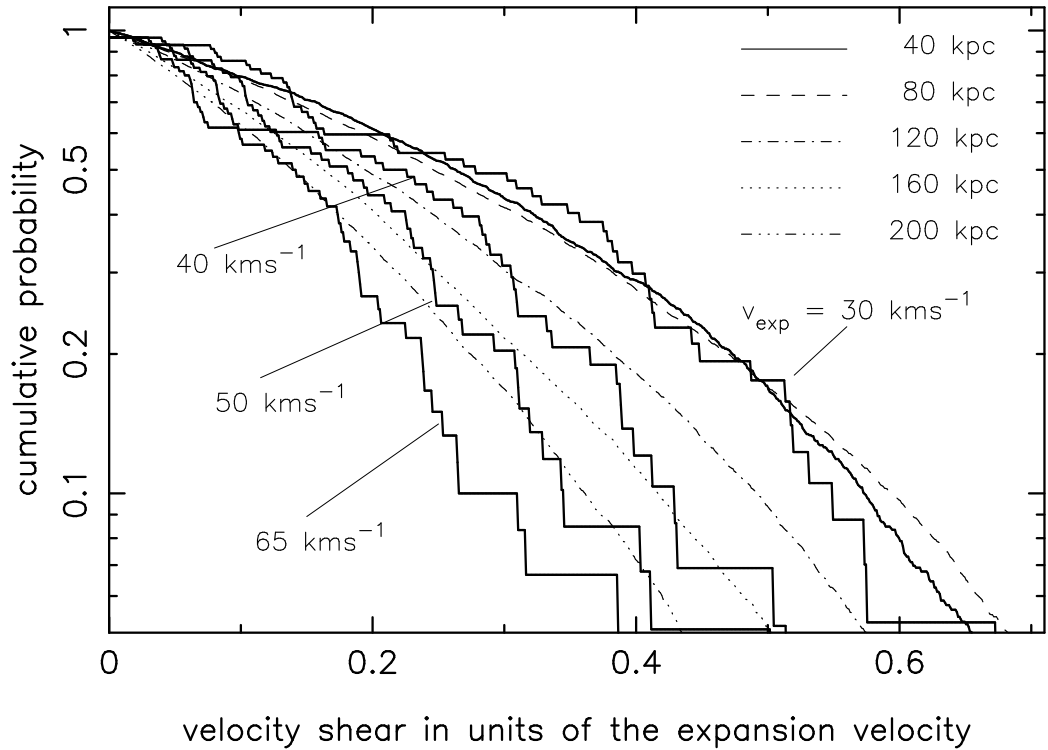}
\caption{\small Cumulative distribution of velocity differences (in units
of the radial expansion velocity $v_{exp}$ for bubble models with proper radii
40, 80, 120, 160 and 200  kpc.
The thick solid lines show the actually observed distribution (see also fig.
\ref{q2345_velewmodels}) scaled along the x-axis to match the model distributions for expansion velocities
30, 40 50, and 65 kms$^{-1}$.\label{cumul}}
\end{figure}

Going a step further, we can compare the {\it shape} of the actually observed distribution of velocity
differences from \weed\ to the hypothetical ones for expanding 
bubbles with different parameter combinations. Fig. \ref{cumul} shows 
the cumulative probability distributions for the observed and 
simulated velocity differences versus the velocity differences
in units of the expansion velocity. The thin lines dropping smoothly
to larger velocities are the models (comprising a single population
with fixed radius and expansion velocity; the radii are given in the
top right hand corner of the plot), and the ragged histograms are the
observed distribution of $\Delta v$. There is obviously only
one observed distribution, which, however, can be modeled either
as arising in a population of small bubbles (in which case the
velocity differences would be a relatively large fraction of the
expansion velocity), or as arising from larger bubbles (where the
expansion velocity would have to be larger and the velocity differences
would constitute a smaller fraction of the expansion velocity).
The measurable velocity difference is linear in the
expansion velocity, so we can scale the observed cumulative
distribution until it matches best a particular combination of radius and 
expansion velocity. It can be seen that reasonable matches 
can be produced between either the bulk of the distributions
or their respective wings, for radius-velocity combinations similar to the ones discussed
in connection with the previous figure, but a single population of bubbles
is not a good match. We cannot proceed any further here without explicitly assuming a distribution
of windshell parameters, which is beyond the scope of this paper. The discrepancy  could be either because a more realistic  distribution of bubble
radii and velocity is required, or because the velocity structure of the Ly$\alpha$
forest has nothing to do with expanding bubbles.

If Ly$\alpha$ clouds were indeed bubble walls or at least owed
their motions to winds, we can get a crude upper limit on the energetics of these winds. 

Assuming that the winds follow a simple expanding shell model like the
one discussed by Mac Low \& McCray (1988), the knowledge of 
the radius of the bubble $R$ and the expansion velocity $v_{exp}$
gives a constraint on the "strength" $L_{38}/n_{-5}$ of the wind:

\begin{eqnarray}
\left(\frac{v_{exp}^3}{157{\mathrm kms}^{-1}}\right)\left(\frac{R^2}{2670{\mathrm pc}}\right) = \frac{L_{38}}{n_{-5}}
\end{eqnarray}

Here $L_{38}$ is the mechanical luminosity (in units of $10^{38}$ erg) and $n_{-5}$ is the particle
number density of the surrounding IGM (in units of $10^{-5}$cm$^{-3}$), assumed to be homogeneous.

A bubble with approximate radius 125 kpc and expansion velocity 55 kms$^{-1}$ compatible with fig.\ref{avvexpr}
would thus have a strength of only $L_{38}/n_{-5}$$\sim 94$, i.e.,
a hundreth of what would be required if winds from Lyman break galaxies
were getting out as far as postulated. Even assuming the largest
radii shown in the diagram, 230 kpc, and velocities consistent
with the observations the strength of the wind
falls short by an order of magnitude. 
The model applied here is of course
hopelessly naive, but more realistic assumptions can only make the discrepancy worse. Assumptions of spherical geometry aside, the energy requirements to get a wind bubble out to a certain radius are certainly
much more exacting when density gradients, infall, and the need to
propel entrained matter are included. Moreover, our estimate for the maximum velocities admitted
is of course conservatively high, as we assumed that all the velocity shear of 16.6 kms$^{-1}$ arises in
winds, and nothing in the Hubble flow or through gravitational motions.

\subsection{Where are the Winds ?}

If our assumptions about the detectability  of winds are correct, then we are led to conclude
that winds by the time we observe them are either too weak or too rare to make an impact on the general IGM.

If high-redshift galactic winds are to be common enough to upset
the gravitational instability picture of
the Ly$\alpha$ forest and be consistent with our observations, the winds must be  rather "limp" or "tired", quite different from
the hundreds of kms$^{-1}$ expansion velocities seen in Lyman break galaxy outflows or in the
component structure of OVI absorbers.
Recent theoretical work (Madau et al 2001; Furlanetto \& Loeb 2003; Fujita et al 2004;
Bertone et al 2005) indicates that the inclusion of infall and entrained matter 
may slow down galactic winds considerably from the hundreds of kms$^{-1}$ seen directly
in the cores of starbursting galaxies to velocities on the order of a few tens of kms$^{-1}$, values consistent with our upper limits
of 45 - 85 kms$^{-1}$ (fig.\ref{avvexpr}), so the observed velocity range in itself is
not a problem.
Are we then seeing wind-driven gas in the
Ly$\alpha$ forest ? The answer is, most probably not. As seen above,
the Hubble expansion plus gravitational collapse does already explain all the observed
velocity shear well, at three different redshifts and separations ranging from sub-kiloparsec 
to 300 kpc scales. If the observed velocity dispersion were dominated by winds,
one would have to explain why the Hubble and gravitational motions are irrelevant and
how the winds conspire to mimic exactly the velocity field in a $\Lambda$CDM universe
without any feedback.

Nevertheless, a generation of old and possibly very widespread winds, perhaps 
connected to the reionization process and to an early phase of heavy-element production, need not
be inconsistent with our observations, if  the residual velocities are smaller
than the limits given here and if pressure equilibrium is able to erase the column density differences between 
the lines of sight. 
The observed very early metal enrichment (Songaila 2001; Pettini et al  2003),
its relative uniformity (Aguirre et al 2005),
the observed mass-metallicity relation (Tremonti et al 2004), and the theoretical difficulties of
getting metals out of massive galaxies (Scannapieco et al 2002; Furlanetto \& Loeb 2003; Fujita et al 2004; Scannapieco 2005) 
all appear to favor an abundance of dwarf galaxies venting their metal-enriched gas early on.
Pushing the hydrodynamic disturbances associated
with the metal enrichment to an early epoch, the close resemblance
of the properties of the Ly$\alpha$ forest to the predictions of a hierarchical scenario can be more
easily reconciled with 
the relatively widespread metal enrichment observed. For example, if winds carried metal-enriched gas to the outer edge of filaments (say to radii of 40 kpc proper) and ceased shortly
after the epoch of reionization (after $z\sim 6$), there would be enough time (1.2 Gyr) until the
redshift of observation ($z\sim 3$) for the
gas to have slipped back into the unaltered CDM potential wells, even at subsonic speeds.

Alternatively, strong winds active at the epoch that we observe (including but not limited to superwinds from Lyman break galaxies) may also be consistent with our observation if the filling factor of winds is small enough to not impact the IGM significantly.
In the absence of realistic wind models it is difficult to use observations to constrain the filling factor
of winds (for an attempt, see Rauch et al 2001b), 
but there are some independent pieces of evidence. If, as Simcoe et al (2002) have suggested, the strong OVI absorbers in their
survey are experimentally identified with galactic wind bubbles from Lyman break galaxies, we have approximately
12 OVI systems at 40\% completeness over a redshift distance $dX=6.9$, or 4.3 systems per
unit redshift. Over the same redshift distance there are about 132 low column density
$(10^{12.5} < N <  10^{14}$ cm$^{-2}$) ordinary Ly$\alpha$ absorption systems at $z\sim2.1$ (Kim et al 2002).
Thus, the relative rate of incidence of wind bubbles to intersections with the general
cosmic web would be about 3\%. This would be the fraction of the {\it volume producing the Ly$\alpha$ forest} that is occupied by winds. It is still possible that winds fill a larger {\it cosmic volume}  if they are
collimated (e.g., DeYoung \& Heckman 1994; Theuns et al 2002) and are preferentially blowing perpendicular to the filaments into the voids. The density gradients into the voids would ease the directional expansion
of the hot gas but would also make detection of this gas with any method very hard. On the other
hand, if such winds were limited to the same structures causing the Ly$\alpha$ forest  they could also occupy an even smaller {\it cosmic volume} than the 3\% of
the cosmic web, in particular if they are strongly clustered. 
Theuns et al (2002), Pieri \& Haehnelt (2004), and Desjacques et al (2004),  
attempting to reproduce the CIV metal distribution, the observed incidence of weak OVI, and the sizes of the Adelberger et al (2003) bubbles, respectively,  have argued
that the cosmic volume filling factor of Lyman break winds is likely to be only on the order
of  few percent. Figuratively speaking, they are just storms in intergalactic teacups. Disturbances that rare would not have affected the velocity distributions discussed above
above, no matter how important their local impact.

\subsection{Alternative Explanations: Winds  or Gravitational Motions ?}

The results discussed here constrain the impact of winds on the IGM, 
but they do not rule out their existence. The original arguments for the existence of high-redshift
winds (large velocity
shifts between emission and absorption lines, possible production sites for the bulk of metals) are certainly persuasive, but the evidence often quoted as proving the impact of winds on the IGM
appears more ambiguous. It is worth speculating whether some of the evidence proposed
in favor of 
superwinds escaping from $z\sim 3$ galaxies does not admit alternative interpretations. 

\smallskip

Adelberger et al (2003) originally suggested that large zones with relatively little HI absorption 
near $z\sim 3$ Lyman break galaxies are the consequence of winds evacuating neutral hydrogen
within radii on the order of 125 kpc (proper). While this result has proven hard to explain
theoretically with any astrophysical effect, the new, larger data set presented by Adelberger et al (2005)
proposes smaller radii (40 kpc) for the average evacuated superwind bubble. We note here that this is
essentially the same size derived by Simcoe et al (2002) for strong OVI absorbers at similar
redshifts, under the assumption that
each absorber is a bubble of highly ionized gas around a Lyman break galaxy. While
there seems to be agreement about the size of the effect, the origin of these regions remains less clear. 
  
Gas heated by compression during gravitational collapse would appear similarly as a halo with a low
fraction of neutral HI gas. In fact, unlike winds, gravitational heating {\it must } take place at some
stage during the formation of every galaxy,
especially in hierarchical structure formation where protogalaxies accrete gas while frequently merging with supersonic
velocities, shocking the ambient gas.  
Judging from 
the  analysis of SPH simulations of forming galaxies
(e.g.,  Rauch et al 1997a, fig. 1), such hot halos or shocked shells with temperatures of several times $10^5 K$ are 
common even around individual merging galaxies at $z\sim3$, with radii of 30-60 kpc proper.  By $z\sim 2$
gaseous halos with temperatures up to $10^6 K$  start engulfing entire groups of these
protogalactic clumps, and hotter,
more spherical large halos with an extent on the order  of 50 -- 100
kpc form quickly around massive galaxies within times on the order of $10^9$ yr.
To explain the factor of 7 decrease in optical depth at the centers of the HI-poor bubbles 
observed by Adelberger et al 2003 by increased thermal ionization would require a rise in temperature by only 1 order
of magnitude (e.g., from $10^4$ to $10^5 K$, for gas overdense by a factor of 10; and less 
for less dense gas; e.g., Haehnelt et al 1996, fig.2),
which is obviously well within what gravitational heating can do.
In the simulation, the evolution to a more spherical, larger hot halo
is rapid (essentially the constituents of a future galaxy
are in free fall), with  hot halos becoming a common feature
below redshift 2, and becoming more common and larger as time proceeds.
Keres et al (2004) and  Birnboim  \& Dekel (2003) 
discuss bimodal galaxy formation
in which part of the galaxy population is fed by accreting
gas with instant cooling, avoiding shock heating during 
infall, whereas another subpopulation grows by
the more orthodox, shocked infall of gas. In any of these scenarios the fraction of galaxies with hot
accretion must be increasing with time, which may provide an observationally testable
prediction. Hot gas halos
are also a basic ingredient in analytical models where
cool gas is fed to a growing galaxy in a multiphase
thermal instability (e.g., Mo \& Miralda-Escud\'e 1996; 
Maller \& Bullock 2004).

It appears that the partial destruction of galaxies in the hierarchical structure formation scenario would also lead
to enhanced IGM metallicities, as observed in the immediate, high-density vicinity of  galaxies
(e.g., Simcoe et al 2002, 2005).
Gnedin (1998) has argued that mergers,
through collision, tidal interactions, and ram-pressure stripping, may be responsible for 
part of the IGM metal enrichment.

Other arguments for the impact of $z\sim 3$ Lyman break galaxies on the IGM have included the
clustering of CIV systems around Lyman break galaxies, which, however, is only indicative
of spatial assocation of both the metals and the galaxies
with the same matter overdensities and does not prove a causal connection, i.e.,
an outflow of the metals out of the same galaxies (e.g., Porciani \& Madau 2005). Given
the large spatial extent of the metals, the latter is quite unlikely
(Scannapieco et al 2005).  
 
We conclude, emphasizing that none of the above rules out high-redshift winds; we only
suggest that their prevalence may be overestimated if the gasdynamical consequences
of the hierarchical merger process are mistaken for winds.

\section{Discussion and Summary}

We have measured the shear between the velocities of absorption systems common to 
close lines of sight to background QSOs. 
Over physical distances on the order of a kiloparsec
the observed distribution of the differences
between the velocities projected along the line of sight is largely consistent with 
being mostly due to measurement error. A small fraction (on the order of 10\% of all systems)
show significant (at the $2.5 \sigma$ level) velocity shear. Inspection of the individual
images shows that the motions mostly appear to be bulk shifts of the entire absorption system
in the two lines of sight.
The mean shift for the 10 largest deviations is 11 kms$^{-1}$, and the rms contribution
to the total width of the distribution of shear is about 6 kms$^{-1}$. We speculate that we may be
seeing rotational or other differential
motion of gas "circling the drain" in a gravitational potential.

Proceeding to larger scales, we measure the velocity shear distribution in the Ly$\alpha$ forest toward three QSO pairs  near mean redshifts 2 and 3.6, for mean separations (60-300 $h_{70}^{-1}$ physical kpc) large enough to see evidence
of the Hubble expansion. The measurement cannot give the absolute value of the Hubble constant,
but only the relative motions of the gas in units of the local Hubble flow.
With increasing separation, the shape of the observed distribution of shear begins to depart from the Gaussian (error-dominated)
shape seen at kiloparsec separations. It shows broad wings as expected if the large-scale systematic
motions take over. Indeed, a simple analytical model
where the absorbers are homologously expanding, randomly oriented pancakes (e.g., Haehnelt 1996) gives a reasonable
representation of the data. Adopting the mean coherence length  from the literature (D'Odorico et al 1998) for the diameter of the pancakes,
the model indicates that the
radial expansion velocity is reasonably close to but somewhat less than the expected Hubble expansion
over that scale. In the case of the lower redshift ($z\sim2$) dataset, 
the best fit indicates that the model pancake would have to expand with $0.8\pm 0.3 (3\sigma)$ of the
local Hubble flow. Confusion (problems with cross-identifying the absorbers
between the lines of sight) is still negligible at this redshift and beam separation, so we can consider
this value as a relatively unbiased measurement, whose main source of error is the finite number of 
absorbers.
For the higher redshift sample the best fit gives a smaller value
($0.65 \times$ the Hubble velocity), but the large confusion involved and doubts about
the validity of our assuming a nonevolving size for the pancakes make us suspect that the  result is a systematic underestimate of the actual expansion velocity. We test these suspicions
with a more sophisticated model using artifical lines of sight to probe the cosmic
web in a cosmological hydrodynamic simulation of a $\Lambda$CDM universe without feedback, with observational parameters as close as possible to the
observed situation. The results from this modeling confirm that the measurement of 
the expansion velocity with the constant-size pancake model applied to an absorption-line sample
selected manually at $z\sim2$ was quite realistic. They further confirm that the  same approach indeed 
underestimates the expansion velocities beyond redshift 3.
A K-S test shows that the observed and simulated distributions of velocity shear are consistent
with being drawn from the same population. The observed rms widths of the
velocity shear distributions,  16.6 kms$^{-1}$ (z=2) and 30.0 kms$^{-1}$ (z=3.6)  closely
resemble the values obtained from the hydrodynamic simulation (14.9 and 30.6 kms$^{-1}$, respectively), and the shapes of the distributions are virtually indistinguishable.
The detailed agreement between the  observed and simulated distributions of velocity shear may be
taken to imply
that whatever physical processes produce the simulated distributions must be present
in reality as well.

We compute the underlying distribution of expansion velocities for {\it absorption-line-selected}
regions in the simulation
(the line-of-sight projection of which produces the distribution of velocity shear).
This distribution shows most Ly$\alpha$ clouds expanding faster than the Hubble flow,
but the mean velocity  (at least at redshift 2 and probably below) is somewhat less
that the Hubble velocity.
The larger fraction of contracting clouds (in comoving coordinates) in the z=2 sample 
as compared to the higher redshift samples may be due in part to deceleration with time or to
the different spatial scales, but it
it could also be partly a selection effect. By imposing an equivalent width detection threshold constant in time, we may be selecting higher density, more collapsed  regions at lower redshift.

The same distribution is also computed for {\it random} regions 
in the simulation. We find significant differences, in that the latter expand increasingly
faster with decreasing redshift than the absorption-line-selected
regions. Apparently, most regions selected by typical Ly$\alpha$ forest absorption lines
show the large-scale kinematics expected of mildly overdense, large sheetlike or
filamentary structures, most of which are draining with super-Hubble velocities into
larger mass agglomerations, while some of them are undergoing  gravitational collapse.

We briefly considered the possibility, occasionally raised, that the Ly$\alpha$
forest could be seriously affected by galactic
feedback, especially galactic superwinds active at the epoch of observation. Given the close agreement between the observed
velocity distribution and the one predicted by the  standard $\Lambda$CDM based gravitational instability 
scenario, we find little room for a cosmological population of superwinds significantly
disturbing the density and velocity structure of the general IGM. While this does not rule
out the existence of such winds, various strands of evidence
suggest that any winds simply may have a small filling factor as far as the overdense IGM
giving rise to the Ly$\alpha$ forest 
is concerned.
To escape detection, high-redshift superwinds may be intrinsically rare, or could be venting preferentially into cosmic voids, or may be more
limited in their individual spatial range and expansion velocity because of the vicissitudes of infall, entrainment, or
the larger ambient density at high redshift.

A more widespread population of {\it early} winds could still be consistent with our measurement and several other recent
constraints on the distribution of metals, as could a later population of "limp" winds with sufficiently low expansion velocities at the time we observe them.

Finally, it appears that much of the observational
evidence usually presented in favor of superwinds in the IGM may not be unique (and may not even
favor superwinds, at least as far as the process of metal enrichment is concerned). Hot halos formed naturally during accretion and mergers in a  hierarchical galaxy formation picture may have  observational properties in common with the HI depleted, metal-enriched bubbles ascribed
to superwinds from massive galaxies. In individual cases, the underlying cause may be hard to ascertain, but
the hierarchical scenario should predict a definite dependence of the radii and rate of incidence
of hot accretion halos with time, which may be tested with observations.

\acknowledgments

MR is grateful to Nick Gnedin for a stimulating discussion that rekindled his interest in 
cross-correlating Ly$\alpha$ forests. He further thanks the NSF for supporting this 
work under grant AST 00-98492. MV and MR thank the Kavli Institute for Theoretical Physics
in Santa Barbara for its hospitality in December 2004 during the workshop
on "Galaxy-Intergalactic Medium Interactions", supported in part by the National Science Foundation under grant No. PHY99-07949. MV thanks PPARC for financial support. MR and GDB thank the I.S. Bowen memorial fund
for superior sustenance.
The work of GDB and WLWS was supported by NSF through grants AST 99-00733 and AST 02-06067.
The hydrodynamic simulation
was done at the UK National Cosmology Supercomputer Center funded by
PPARC, HEFCE and Silicon Graphics/Cray Research.

\pagebreak

\clearpage

\clearpage

\clearpage

\clearpage

\begin{deluxetable}{cccccc}
\small
\tablewidth{0pt}
\tablenum{1}
\tablecaption{Observational Data vs. Simulations\tablenotemark{a}}
\tablehead{
\colhead{absorption line sample} &
\colhead{$\overline{z}$} &
\colhead{$H(z)$} &
\colhead{$\overline{d}$ [$h_{70}^{-1}$kpc]} &
\colhead{$(\Delta\overline{v})_{RMS}(obs.)$} &
\colhead{$(\Delta\overline{v})_{RMS}(sim.)$} 
}
\startdata
\rxj A,B               &2.57& 243.7 & 0.82  & $< 6.3$&\nodata  \\
\weed\                 &2.04& 195.2 & 61.0    & 16.6& 14.9 \tablenotemark{b}\\ 
Q1422/1424 \& Q1439A,B &3.62& 352.8 & 260.5 & 30.0& 30.6 \tablenotemark{c}\\
\enddata
\tablenotetext{a}{\small Note that there are slight differences between the mean redshifts and
the cosmological parameters adopted for the analysis of the data
($\Omega$=0.25, $\Lambda$=0.75, $h$=0.70) and the simulations
($\Omega$=0.26, $\Lambda$=0.74, $h$=0.72).}
\tablenotetext{b}{\small The simulated distribution was obtained for z=2.0.}
\tablenotetext{c}{\small The simulated distribution is the mean from
two simulations done at z=3.4 and z=3.8.}

\end{deluxetable}

\clearpage

\begin{deluxetable}{ccccccc}
\small
\tablecolumns{7} 
\tablewidth{0pt}
\tablenum{2}
\tablecaption{Hubble Ratios in the Simulations}
\tablehead{
\colhead{} &\colhead{} &  \multicolumn{2}{c}{absorption selected regions}    & \colhead{}   &
\multicolumn{2}{c}{random regions} \\ 
\cline{3-4} \cline{6-7}  
\tablevspace{-.3cm}
\colhead{$\overline{z}$} &
\colhead{$d$ [$h_{72}^{-1}$kpc]} &
\colhead{mean $ r$} &
\colhead{median $r$} &
\colhead{}  &
\colhead{mean $r$} &
\colhead{median $r$} 
}
\startdata
2.0  & 61      &0.85& 1.08 && 1.08 &1.22 \\
3.4  & 288     &1.02& 1.09 &  & 1.09 &1.16 \\ 
3.8  & 236     &1.03& 1.11  && 1.09&1.15\\
\enddata
\end{deluxetable}

\end{document}